%% file: ustrong.tex
\documentclass[twocolumn,floatfix]{aastex631}
\usepackage{color,colortbl}

\usepackage{natbib}
\usepackage{longtable}
\usepackage{graphicx}
\usepackage{amsmath}
\usepackage{amssymb}
\usepackage{hyperref}
\usepackage{threeparttable}
\usepackage{tabularx,ragged2e}
\usepackage{booktabs, array, times}
\usepackage{multirow}
\usepackage{epstopdf}
\usepackage{capt-of}
\usepackage{rotating}
\usepackage{subfigure}
\usepackage{collcell}

\DeclareUnicodeCharacter{2212}{-}
\definecolor{comp1}{HTML}{DB1B65}  
\definecolor{comp2}{HTML}{05ADDB}  
\definecolor{comp3}{HTML}{AAC321}  
\definecolor{comp4}{HTML}{34990A}  
\definecolor{comp5}{HTML}{F1FD79}  
\definecolor{comp6}{HTML}{76BE24}  
\definecolor{comp7}{HTML}{6E668E}  
\definecolor{comp8}{HTML}{2C3D7C}  
\definecolor{comp9}{HTML}{F0705E}  
\definecolor{comp10}{HTML}{EB9ACA} 
\definecolor{comp11}{HTML}{B4FBD2} 
\definecolor{comp12}{HTML}{5D2A88} 
\definecolor{comp13}{HTML}{D5B15A} 
\definecolor{comp14}{HTML}{140177} 
\definecolor{comp15}{HTML}{040A38} 
\definecolor{comp16}{HTML}{34F392} 
\definecolor{comp17}{HTML}{590DAD} 
\definecolor{comp18}{HTML}{42ED51} 
\definecolor{comp19}{HTML}{002669} 
\definecolor{comp20}{HTML}{B71F9E} 
\definecolor{comp21}{HTML}{FDCF1C} 
\definecolor{comp22}{HTML}{3F8310} 
\definecolor{comp23}{HTML}{D877DC} 
\definecolor{comp24}{HTML}{537084} 
\definecolor{comp25}{HTML}{E513DF} 
\definecolor{comp26}{HTML}{22DBC7} 
\definecolor{comp27}{HTML}{3393DA}
\def\CIVdblt{{\rm C}\kern 0.1em{\sc iv}~$\lambda\lambda 1548, 1550$}
\def\NVdblt{{\rm N}\kern 0.1em{\sc v}~$\lambda\lambda 1238, 1242$}
\def\OVIdblt{{\rm O}\kern 0.1em{\sc vi}~$ 1031, 1037$}
\def\SVIdblt{{\rm S~}\kern 0.1em{\sc vi}~$ 933, 944$}
\def\SiIVdblt{{\rm Si~}\kern 0.1em{\sc iv}~$\lambda\lambda 1394, 1403$}
\def\MgIIdblt{{\rm Mg}\kern 0.1em{\sc ii}~$\lambda\lambda 2796, 2803$ }
\def\AlIIIdblt{{\rm Al}\kern 0.1em{\sc iii}~$\lambda\lambda 1854, 1862$}
\def\NeVIIIdblt{{\rm Ne~}\kern 0.1em{\sc viii}~$\lambda\lambda770, 780$}
\def\NeV{\hbox{{\rm Ne~}\kern 0.1em{\sc v}}}
\def\NeVI{\hbox{{\rm Ne~}\kern 0.1em{\sc vi}}}
\def\NeVIII{\hbox{{\rm Ne~}\kern 0.1em{\sc viii}}}
\def\OII{\hbox{{\rm O}\kern 0.1em{\sc ii}}}
\def\OIII{\hbox{{\rm O~}\kern 0.1em{\sc iii}}}
\def\OIV{\hbox{{\rm O~}\kern 0.1em{\sc iv}}}
\def\OV{\hbox{{\rm O~}\kern 0.1em{\sc v}}}
\def\OVI{\hbox{{\rm O}\kern 0.1em{\sc vi}}}
\def\OVII{\hbox{{\rm O~}\kern 0.1em{\sc vii}}}
\def\OVIII{\hbox{{\rm O~}\kern 0.1em{\sc viii}}}
\def\NI{\hbox{{\rm N~}\kern 0.1em{\sc i}}}
\def\NII{\hbox{{\rm N~}\kern 0.1em{\sc ii}}}
\def\NIII{\hbox{{\rm N~}\kern 0.1em{\sc iii}}}
\def\NIV{\hbox{{\rm N~}\kern 0.1em{\sc iv}}}
\def\NV{\hbox{{\rm N~}\kern 0.1em{\sc v}}}
\def\NVII{\hbox{{\rm N~}\kern 0.1em{\sc vii}}}
\def\CII{\hbox{{\rm C}\kern 0.1em{\sc ii}}}
\def\CIII{\hbox{{\rm C}\kern 0.1em{\sc iii}}}
\def\SiII{\hbox{{\rm Si}\kern 0.1em{\sc ii}}}
\def\SiIII{\hbox{{\rm Si}\kern 0.1em{\sc iii}}}
\def\SiIV{\hbox{{\rm Si}\kern 0.1em{\sc iv}}}
\def\SIV{\hbox{{\rm S~}\kern 0.1em{\sc iv}}}
\def\SV{\hbox{{\rm S~}\kern 0.1em{\sc v}}}
\def\SVI{\hbox{{\rm S~}\kern 0.1em{\sc vi}}}
\def\SiI{\hbox{{\rm Si~}\kern 0.1em{\sc i}}}
\def\PII{\hbox{{\rm P~}\kern 0.1em{\sc ii}}}
\def\AlI{\hbox{{\rm Al~}\kern 0.1em{\sc i}}}
\def\AlII{\hbox{{\rm Al~}\kern 0.1em{\sc ii}}}
\def\AlIII{\hbox{{\rm Al~}\kern 0.1em{\sc iii}}}
\def\CaI{\hbox{{\rm Ca}\kern 0.1em{\sc i}}}
\def\CaII{\hbox{{\rm Ca}\kern 0.1em{\sc ii}}}
\def\CrII{\hbox{{\rm Cr~}\kern 0.1em{\sc ii}}}
\def\CII{\hbox{{\rm C}\kern 0.1em{\sc ii}}}
\def\CIII{\hbox{{\rm C}\kern 0.1em{\sc iii}}}
\def\CIV{\hbox{{\rm C}\kern 0.1em{\sc iv}}}
\def\CV{\hbox{{\rm C}\kern 0.1em{\sc v}}}
\def\MgX{\hbox{{\rm Mg}\kern 0.1em{\sc x}}}
\def\MgI{\hbox{{\rm Mg}\kern 0.1em{\sc i}}}
\def\MgII{\hbox{{\rm Mg}\kern 0.1em{\sc ii}}}
\def\FeII{\hbox{{\rm Fe}\kern 0.1em{\sc ii}}}
\def\FeIII{\hbox{{\rm Fe~}\kern 0.1em{\sc iii}}}
\def\TiII{\hbox{{\rm Ti}\kern 0.1em{\sc ii}}}
\def\NaI{\hbox{{\rm Na~}\kern 0.1em{\sc i}}}
\def\SII{\hbox{{\rm S}\kern 0.1em{\sc ii}}}
\def\H{\hbox{{\rm H~}}}
\def\HI{\hbox{{\rm H}\kern 0.1em{\sc i}}}
\def\HeI{\hbox{{\rm He~}\kern 0.1em{\sc i}}}
\def\HII{\hbox{{\rm H}\kern 0.1em{\sc ii}}}
\def\Lya{\hbox{{\rm Ly}\kern 0.1em$\alpha$}}
\def\Lyb{\hbox{{\rm Ly}\kern 0.1em$\beta$}}
\def\Lyg{\hbox{{\rm Ly}\kern 0.1em$\gamma$}}
\def\Lyth{\hbox{{\rm Ly}\kern 0.1em$\theta$}}
\def\Lyfive{\hbox{{\rm Ly}\kern 0.1em$5$}}
\def\Lysix{\hbox{{\rm Ly}\kern 0.1em$6$}}
\def\Lyseven{\hbox{{\rm Ly}\kern 0.1em$7$}}
\def\Lyeight{\hbox{{\rm Ly}\kern 0.1em$8$}}
\def\Lynine{\hbox{{\rm Ly}\kern 0.1em$9$}}
\def\Lyten{\hbox{{\rm Ly}\kern 0.1em$10$}}
\def\MnII{\hbox{{\rm Mn}\kern 0.1em{\sc ii}}}
\def\kms{\hbox{km~s$^{-1}$}}
\def\cmsq{\hbox{cm$^{-2}$}}
\definecolor{LightCyan}{rgb}{0.88,1,1}
\newcommand{\hi}{\mbox{H\,{\sc i}}}
\def\cc{\hbox{cm$^{-3}$}}
\newcommand{\angstrom}{\mbox{\normalfont\AA}}
\newcommand{\CLOUDY}{\textsc{cloudy}}
\newcommand{\hdenu}{$\log (n_{\rm H}/cm^3)$}
\newcommand{\coldenu}{$\log [N(\hi)/cm^2]$}
\newcommand{\thickness}{$\log L$}
\newcommand{\tempu}{$\log (T/K)$}
\newcommand{\metallicity}{$\log (Z/Z_{\sun})$}
\newcommand{\hden}{$\log (n_{\rm H})$}
\newcommand{\colden}{$\log [N(\hi)]$}

\newcommand{\btherm}{b$_{\rm {T}} (\rm {\hi})$}
\newcommand{\bturb}{b$_{n\rm {T}}$}
\newcommand{\bnet}{b$(\rm {\hi})$}

\begin{document}

\title{Kinematic analysis of an Ultra-Strong {\MgII} absorber at $z \approx 1.13$ linking to Circumgalactic Gas Structures}


\author[0009-0007-3947-5838]{Purvi Udhwani}
\affiliation{Department of Earth and Space Sciences, Indian Institute of Space Science \& Technology, Thiruvananthapuram 695547, Kerala, INDIA}

\author[0000-0001-9966-6790]{Sameer}
\affiliation{Department of Astronomy \& Astrophysics, The Pennsylvania State University, 525 Davey Laboratory
University Park, PA, 16802, USA}
\affiliation{Department of Physics \& Astronomy, University of Notre Dame, Notre Dame, IN 46556 USA}

\author[0000-0002-1490-5367]{Anand Narayanan}
\affiliation{Department of Earth and Space Sciences, Indian Institute of Space Science \& Technology, Thiruvananthapuram 695547, Kerala, INDIA}


\author[0000-0003-3938-8762]{Sowgat Muzahid}
\affiliation{Inter-University Centre for Astronomy \& Astrophysics, Post Bag 4, Ganeshkhind, Savitribai Phule Pune University Campus, Pune 411 007, India}

\author[0000-0003-4877-9116]{Jane Charlton}
\affiliation{Department of Astronomy \& Astrophysics, The Pennsylvania State University, 525 Davey Laboratory
University Park, PA, 16802, USA}

\author[0000-0001-5804-1428]{Sebastiano Cantalupo}
\affiliation{Astrophysics Unit, Department of Physics, University of Milano-Bicocca, Piazza della Scienza 3, 20126 Milan, Italy.}
\shorttitle{Ultrastrong MgII}
\shortauthors{Udhwani et al.}

\begin{abstract}
We present a spectroscopic and imaging analysis of the $z_{gal} \approx 1.1334$ ultra-strong {\MgII} absorption system identified in the $VLT$/UVES spectrum of a background quasar located at $\rho \approx 18$~kpc from a star-forming galaxy. Low ionization metal lines like {\MgI}, {\FeII}, and {\CaII} are also detected for this absorber. The $\HI$ lines are outside of the wavelength coverage. The {\MgII} has a rest-frame equivalent width of $W_r(2796) =3.185\pm0.032$~{\AA}, with the absorption spread across $\Delta v \approx 460$~{\kms} in several components. A component-by-component ionization modeling shows several of these components having solar and higher metallicities. The models also predict a total {\HI} column density of $\log~[N(\HI)/\cmsq] \approx 22.5$, consistent with ultra-strong {\MgII} absorbers being sub-Damped Lyman Alpha and Damped Lyman Alpha systems. The absorber is well within the virial radius of the nearest galaxy which has a stellar mass $M_* = 4.7 \times 10^{10}$~M$_\odot$, and a star formation rate of $\approx  8.3$~M$_\odot$~yr$^{-1}$. The absorption is along the projected major axis of the galaxy with a velocity spread that is wider than the galaxy's disk rotation. From the kinematic analysis of the absorber and the galaxy, the origin of the absorption can be attributed to a combination of circumgalactic gas structures, some corotating with the disk and the rest at line-of-sight velocities outside of the disk rotation.

\end{abstract}



\section{Introduction} \label{sec:intro}

The extended envelope of gas surrounding galaxies, referred to as the circumgalactic medium (CGM), is a cache of the galaxy's stellar and dynamical activities tracing gas accretion from the intergalactic medium (IGM), outflows driven by active galactic nuclei (AGN) and supernovae, and tidally displaced interstellar gas \citep{2017ARA&A..55..389T,2014ApJ...786...54P,2011Sci...334..948T,2012ARA&A..50..491P}. Cumulatively, these baryon recycling processes are understood to play a central role in the evolution of galaxies. Absorption line studies using background quasars as probes are one of the most sensitive ways to trace the halo gas, especially of low densities, out to galactocentric distances of several hundred kilo-parsecs (e.g.: \citealt{1969ApJ...156L..63B,1995AAS...186.2507L,2021AAS...23731301W}). Such studies provide line-of-sight information on column densities, chemical composition, density-temperature structure, and kinematics of gas moving in and out of galaxies, which are not primarily traceable through emission line observations. Different ionic species, usually resonant doublets, are used as diagnostics for probing the complex multi-phase structure of the CGM gas. Insights on the highly ionized hot ($T \gtrsim 10^5$~K) component of the halo have come predominantly from observations of {\OVI} absorption, whereas {\CIV}, {\SiIV}, along with {\MgII}, {\CII}, and {\SiII} have served as tracers of the cooler ($T \lesssim 10^{4-5}$~K) multiphase photoionized halo gas, with the low ions often arising in compact parsec scale, denser regions embedded within the more diffuse halo \citep{2000,2004ApJS..155..351S,2006hst..prop10925S,2006ApJ...643L..77T,2010ApJ...721..960N,2013ApJ...776..115N,2014ApJ...784....5M,2019MNRAS.485.1961Z,Lehner_2019,2021MNRAS.503.3243A}. 

The {\MgIIdblt} resonant doublet is routinely used to study absorption from the gaseous halos and the disks of intervening spiral galaxies. The rest-wavelength of the doublet transitions allows one to detect such systems using ground-based optical spectroscopic observations at moderately low redshifts where follow-up deep galaxy observations are feasible. Such studies have found several strong {\MgII} absorbers [equivalent width in the absorber frame, $W_r$ (\MgII~2796~\AA) $\gtrsim 0.3$~{\AA}] to be coincident with ($\sim 40h^{-1}$~kpc of projected separation) luminous galaxies ($L_K > 0.06 L^{*}_{K}$) \citep{article,1991A&A...243..344B,1992A&A...257..417B,1993A&A...279...33L,articles,2005pgqa.conf...24C,10.1111/j.1365-2966.2007.11740.x}. 

A recurring trend in these absorbers-galaxy studies is the anti-correlation between the strength of the {\MgII} absorption feature ($W_r$) and the impact parameter ($\rho$) to the closest galaxy. Initial studies that established this association were primarily based on identifying galaxies at small impact parameters to the quasar line of sight, with galaxy systemic redshifts coinciding with the absorber \citep{articles,2010ApJ...724L.176C, 2013ApJ...776..115N}. The kinematics of the absorbing gas as seen in the {\MgIIdblt} profiles has been modeled as a combination of absorption from gas co-rotating with the thick disk of the galaxy \citep{Steidel_2002}, and halo clouds moving with a radial component of velocity \citep{Charlton}. Subsequent {\MgII} absorption line surveys reinforced the galaxy connection through trends between the absorbing gas and the nearest galaxy properties. Using a sample of more than 8500 strong {\MgII} absorbers from Sloan Digital Sky Survey (SDSS) quasar spectra, \citet{2011MNRAS.417..801M} found a $\sim\,15\sigma$ correlation between the rest-frame equivalent width of the {\MgII} doublet lines and the nearest galaxy star formation rates estimated from its [\OII] luminosity.  Compared to passive red galaxies, a higher covering fraction of low-ionization gas traced by {\MgII} for star-forming galaxies was seen in other studies as well \citep{2012ApJ...760L...7K,2013ApJ...776..115N,2014ApJ...795...31L,2018ApJ...866...36L}. Such trends were also seen in down-the-barrel spectroscopic studies in the form of strong {\MgII} absorption blue-shifted relative to the rest-frame of star-forming galaxies \citep{2009AIPC.1201..142W,2010ApJ...719.1503R}. These signify feedback from star formation as one of the factors contributing to the {\MgII} absorption cross-section around galaxies. In a broader cosmological context, this trend is reflected in the co-moving redshift number density of strong {\MgII} absorbers ($0.3 \leq W_r(2796) \leq 1$~{\AA}) closely following the peak in global star formation density at $z \approx 2$, and its subsequent decline \citep{2017ApJ...850..188C,2000} towards lower redshifts. 

Absorption line studies targeting the CGM of foreground galaxies have revealed other interesting trends between the strength of low-ionization absorption, and the orientation of galaxies relative to the line of sight. A recurring finding has been the bimodality in the distribution of {\MgII} covering fraction with respect to the azimuthal angle\footnote{Azimuthal angle is the angle between the QSO line of sight and the projected galaxy major axis.} of the quasar line of sight \citep{2012MNRAS.426..801B,2012ApJ...760L...7K}. The covering fraction of {\MgII} tends to be large ($f_c \gtrsim 80\%$) when the line of sight is not far from the galaxy's projected major or minor axis, as opposed to intermediate azimuthal orientations \citep{2012ApJ...760L...7K}. Such a bimodality in the distribution of {\MgII} against azimuthal angle is seen across different absorber-galaxy samples (e.g., \citealt{2011ApJ...743...10B}). The overall trend in covering fraction is attributed to the CGM along the extensions of the galaxy minor axis populated by biconical winds which are preferentially directed along the minor axis of galaxies, as well as pristine or recycled gas that gets accreted along the extended major axis within the plane of the galactic disk.(e.g.,  \citealt{2015ApJ...812...83N}, \citealt{2017gefb.confE..20H}, \citealt{2019MNRAS.485.1961Z}). Such accreted gas extends the gas cross-section of the disk to distances of $\approx 70$~kpc, tends to co-rotate with the disk, and is, therefore, a means to study disk kinematics far beyond the emitting gas \citep{2019MNRAS.490.4368S}. The {\MgII} has also been detected in emission in the CGM of star-forming galaxies (e.g., \citealt{2011ApJ...728...55R,2013ApJ...770...41M}). The spatial extent of the emission is often consistent with tracing wind material in a biconical outflow \citep{Guo2023, Zabl_2021} although there are cases where no specific directionality is seen for the emission \citep{2021ApJ...909..151B}.

The environment also plays an important role in the widespread distribution of metals around galaxies. \citet{2020MNRAS.499.5022D} find in their galaxy-absorber study that nearly two-thirds of the {\MgII} systems in their sample of 21 have two or more galaxies within $\Delta v \sim 500$~{\kms} and $\rho \sim 350$~kpc of the absorber. In such cases, the {\MgII} absorption is found to be significantly stronger compared to the CGM of isolated galaxies. Similar results were also found by \citet{2018ApJ...869..153N} where the equivalent width and covering fraction ($f_c \approx 0.9$) of the {\MgII} absorbing gas was higher in group environments compared to field galaxies. The enhanced spatial distribution of cold gas in group environments is attributed to frequent gravitational interactions between galaxies and their dwarf satellites that result in the displacement of interstellar gas to the surrounding environment. Whether the {\MgII} profile in such cases is a result of absorption from multiple halos blended, or from gas belonging to the overall group environment remains ambiguous \citep{2020MNRAS.499.5022D,2011ApJ...743...10B}. The clustering of {\MgII} gas near Luminous Red Galaxies suggests that tidal interactions and ram-pressure stripping can enhance the covering fraction of cold gas in halos hosting even quiescent galaxies \citep{2006MNRAS.371..495B,2009ApJ...698..819L,2009ApJ...702...50G,2014AAS...22313702Z,2016yCat..18210114H}. 

Modeling the physical conditions and kinematics of strong {\MgII} offers a way to capture these diverse processes that regulate the baryon cycle in galaxies. Absorbers with unusually strong {\MgII} with $W_r(\MgII~\lambda2796) \gtrsim 3$~{\AA} and large velocity spreads of $|\Delta v| \approx 300 - 600$~{\kms} are generally referred to as ultrastrong {\MgII} systems \citep{2007ApJ...658..185N,2007hsct.confE..55Z}. They are identified at low impact parameters to galaxies ($\rho < 40$~kpc), with origins in galactic winds \citep{2011MNRAS.412.1559N} and/or large-scale matter accretion onto the disk \citep{2011ApJ...738...39S}. These ultrastrong absorbers thus offer a pathway to study the ionization properties and dynamics of such gas within the inner circumgalactic halo, where it is most strongly influenced by the galaxy \citep{2007ApJ...668L.123M}. The redshift number density of the sub-population of these ultrastrong {\MgII} is found to decline strongly with redshift, from $z \approx 2$ to $z \approx 0.5$ \citep{2005ApJ...628..637N}. This trend can be a fallout of the evolution in several galactic scale processes such as star formation, interactions, and mergers between galaxies. The ultrastrong {\MgII} absorber population also shows certain divergent trends from the broader category of {\MgII} systems. The closest galaxies associated with ultrastrong {\MgII} absorbers are found at impact parameters larger than what is anticipated based on the $W_r - \rho$ anti-correlation observed in the general population of {\MgII} systems \citep{2022MNRAS.513.3836G}. For a given impact parameter, the host galaxies of ultrastrong {\MgII} absorbers tend to be more massive and luminous than those associated with standard {\MgII} systems \citep{2022MNRAS.513.3836G}. Additionally, their galaxies have lower SFRs compared to the typical value along the galaxy main sequence, similar to galaxies that are transitioning from star-forming to quiescent systems \citep{2024MNRAS.527.5075G}. 


In this work, we report on the properties of an ultrastrong {\MgII} absorber at $z_{abs} \approx 1.133$ identified in the $VLT$/UVES spectrum towards the background quasar Q~$1621-0042$. The galaxy associated with this absorber was discovered in the MUSEQuBES survey for studying the CGM of {\Lya} emitters at $z > 3$ \citep{10.1093/mnras/staa1347,2021MNRAS.508.5612M}. We compare the properties of the absorbing gas with the galaxy properties and present plausible scenarios for the origin of this ultrastrong {\MgII} system. In Sec~\ref{sec:sec2}, we describe the UVES absorption line data and the $VLT$/MUSE integral field unit (IFU) data and their data reduction procedures. The absorber and galaxy spectral line measurements are presented in Sec~\ref{sec:sec3}.  The physical conditions in the absorbing gas obtained from a component-by-component ionization modeling is explained in Sec~\ref{sec:sec4}. We conclude with a discussion on how the absorbing gas is linked to the CGM of the nearest galaxy. Throughout, we adopt a $\Lambda$CDM cosmology with $\Omega_m=0.27$, $\Omega_b=0.046$ and $H_0=70$ {\kms} Mpc$^{-1}$. All logarithms are to the base of 10. 


\section{Data and Observations}\label{sec:sec2}

\subsection{Spectroscopic Data}
\label{sec:data}
    The $VLT$/UVES optical spectra of the Q~$1621-0042$ quasar [$z_{em} = 3.728$, $(\alpha, \delta) = 245.320458^o, -0.714194^o$] was taken from the reduced, and continuum fitted UVES Spectral Quasar Absorption Database (SQUAD) DR1 \citep{2019MNRAS.482.3458M}. The quasar was observed as part of Prog IDs. $075.A-0464$(A)(PI: T.S.Kim), $091.A-0833$(A)(PI: Joop Schaye), $093.A-0575$(A)(PI: Joop Schaye) on 2005-03-21, for a cumulative exposure time of $t_{exp} = 24.05$~h split into 23 observations. The UVES data has a spectral resolution of $R = \lambda/\Delta \lambda \approx 48440$ over the wavelength range $3280 - 9460$~{\AA}. The spectra were air-to-vacuum corrected and binned to $0.04$~{\AA} corresponding to the Nyquist sampling bin size for the blue part of the spectrum. The binned spectrum has a mean $S/N = 85 - 90$~per resolution element redward of the {\Lya} forest. Lines within the wavelength of $\lambda \lesssim 5750$~{\AA} are associated with the {\Lya} forest. Some of the metal lines associated with the $z_{abs} = 1.13331$ absorber are in the forest region, as we describe later.

\subsection{Integral Field Spectrograph Data}\label{sec:ifudata}

    The field centered on this quasar was observed using the VLT/MUSE-Multi Unit Spectroscopic Explorer \citep{2010SPIE.7735E..08B} under the program ID. 095.A$-$0200(A) (PI: Joop Schaye) for a total exposure time of 9\,h. The galaxy spectra extracted from the MUSE data cube lies within a wavelength range of $4750 - 9350$~{\AA} (air) and has a spectral resolution of $R \sim 3000$. MUSE has a $1^{\prime} \times 1^{\prime}$ field of view in the wide-field mode and with a spatial sampling of $0.2'' \times 0.2''$. This quasar field observation was part of the MUSEQuBES survey and was reduced using MUSE pipeline v1.6 and post-processed (improve flat fielding and sky subtraction) using the tools in the CubExtractor package \citep[CubEx;][]{2019MNRAS.483.5188C}. The details of the MUSEQuBES survey can be found in section~2.1 of \citet{2021MNRAS.508.5612M}.

\begin{figure*} 
    \centering 
        \includegraphics[width=0.45\textwidth]{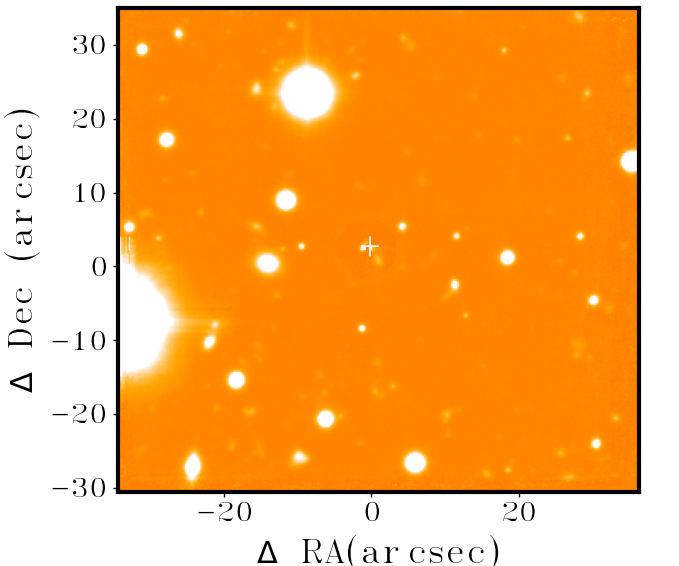}
        \includegraphics[width=0.45\textwidth]{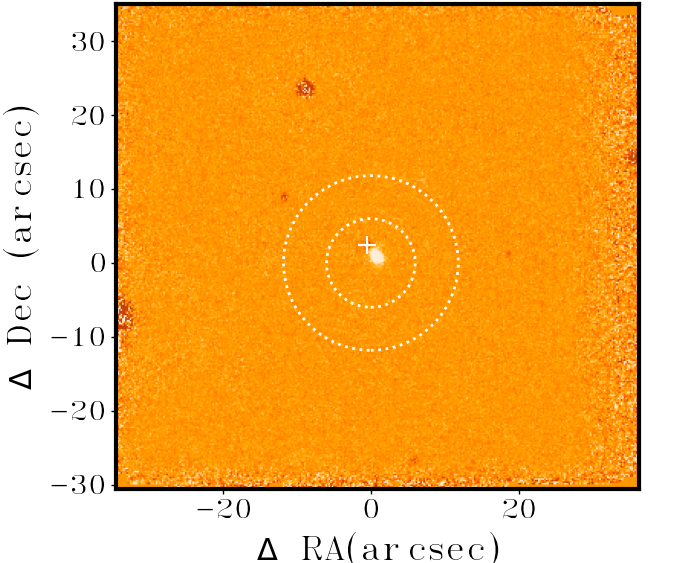}
        \includegraphics[width=\textwidth]{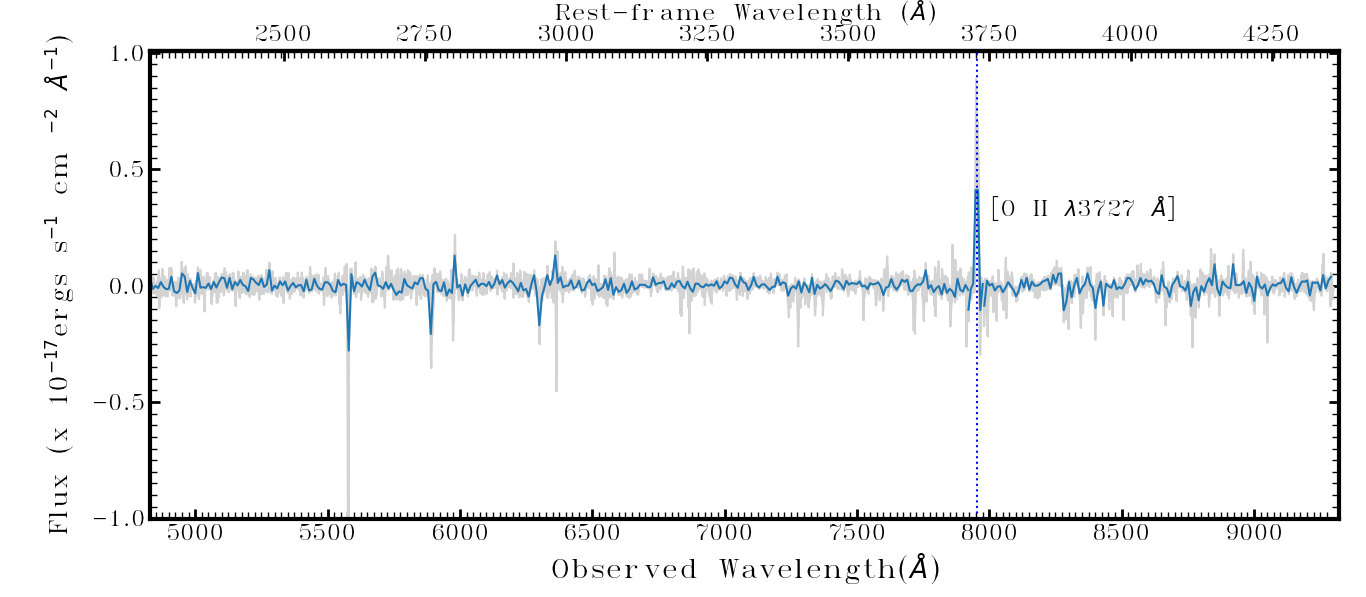}
        \caption{\textit{Top-Left} : White light image from the $VLT$/MUSE with dimensions of $1^{\prime} \times 1^{\prime}$ centered on the quasar Q\,$1621-0042$. The quasar light is removed by subtracting the PSF as indicated in \S\,\ref{sec:ifudata}.  \textit{Top-Right}: A narrow band image of the same field centered around the [\OII]~$\lambda\lambda3726, 3728$~{\AA} doublet at $z_{gal} = 1.1334$. The [\OII] doublet lines are unresolved at the resolution of MUSE. The line of sight to the background quasar Q\,$1621-0042$ is indicated with the cross-wire symbol. The circles with dotted lines represent $\rho = 50$~kpc and $\rho =100$~kpc projected separations around the galaxy at $z_{gal}$ respectively. The foreground field of the quasar does not show any other extended source at the redshift of $z_{gal}\approx$1.1334. \textit{Bottom}: The galaxy spectrum obtained from MUSE is shown in grey. The \textit{blue} curve is the binned spectrum. The redshifted [\OII] emission lines are used to establish the redshift of the galaxy coinciding with the absorber, and determine star-formation rate of the galaxy.}
        \label{white_light}
\end{figure*}

    The MUSE white light and a narrow-band image of the $1^{\prime} \times 1^{\prime}$ field centered around the quasar are shown in Figure~\ref{white_light}. At an impact parameter of $\approx 18$~kpc ($\approx 2.1^{\prime \prime}$), the foreground galaxy coincident with the absorber has a GOTOQ (Galaxy On Top of Quasar) configuration i.e. along the line of sight, the galaxy is positioned in front of the quasar  \citep{10.1111/j.1365-2966.2007.11740.x} The quasar PSF has to be therefore subtracted to separate the galaxy emission from the overlapping quasar emission as seen in the MUSE field. For this, we subtract a wavelength-dependent PSF best-fit model for the quasar component from each spaxel$^*$\footnote{A spaxel is a 3D pixel or spectral pixel where each point also has a third wavelength dimension along with the two spatial dimensions.} around the center except for $< 1^{\prime\prime}$ using the  method as discussed in section 2 of \citet{2018ApJ...869L...1J}. We used the PampelMuse code \citep{2013A&A...549A..71K} to determine the PSF parameters of stars in the field and get an accurate estimate of the seeing value. This was done by fitting a Moffat profile on a few stars in the field and comparing their profiles at different wavelengths to get the best values for the FWHM of the PSF and $\beta$ (wings of the PSF). The spatial resolution is seeing limited, which for this case is estimated to be $0.62^{\prime\prime}$. A linear continuum flux level is subtracted from the cube to take out the non-emission line sources such as stellar objects, and a sub-cube is generated centered around the galaxy coordinates with a fixed radius of a few-arcsecond by cropping the cube along the spatial axes using MPDAF (MUSE Python Data Analysis Framework) Python package \citep{2016ascl.soft11003B}. The galaxy's one-dimensional spectrum is extracted from this resultant sub-cube. 

\section{Absorption line analysis of UVES data}\label{sec:sec3}

\begin{figure*}
    \centering
    \includegraphics[width=\textwidth]{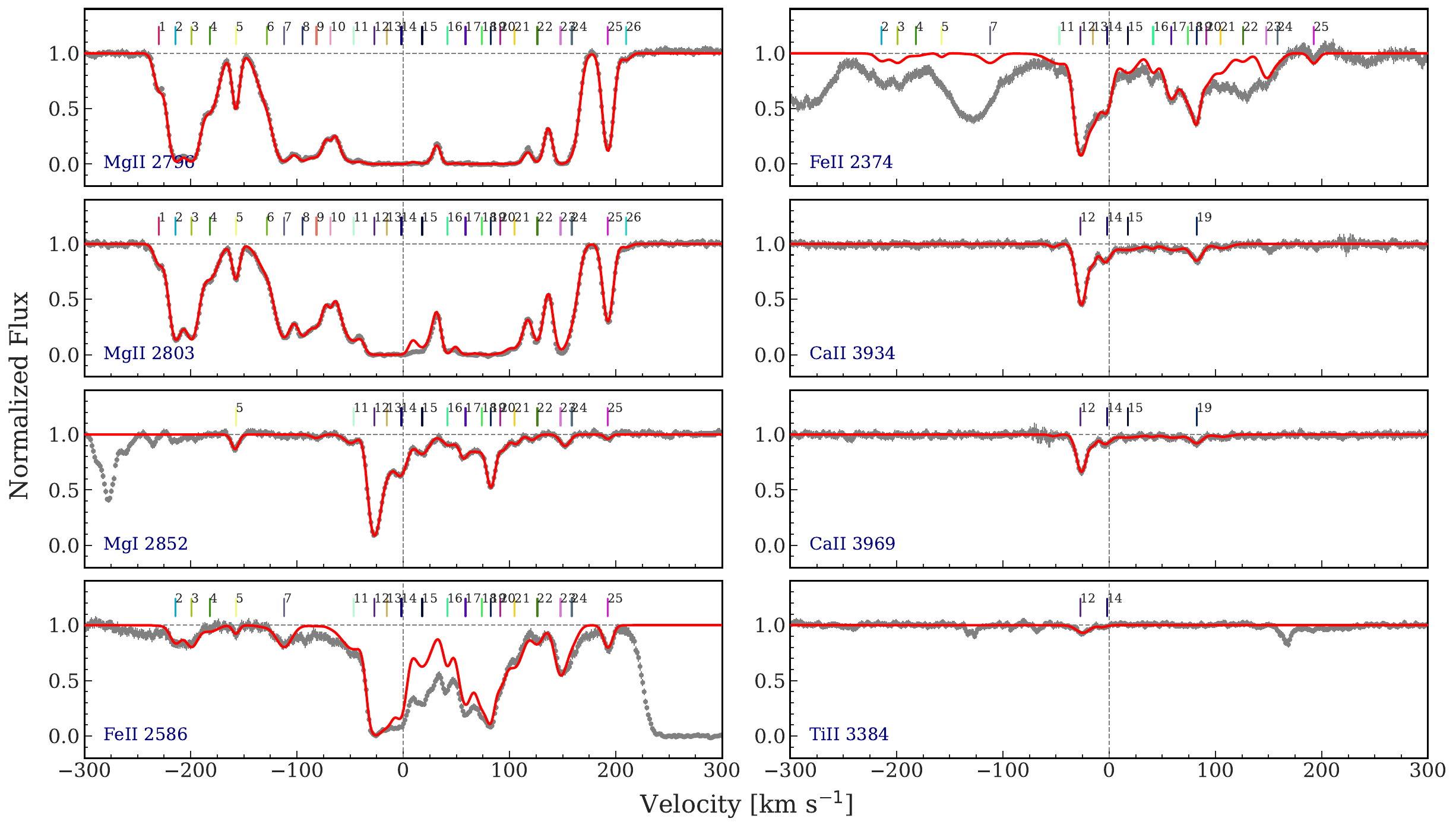}
    \caption{The system plot centered on the emission redshift of z = 1.1334 of the nearest galaxy, showing the lines detected at $>\,3$ sigma and their respective Voigt profile fits (red) imposed over the QSO UVES spectra (grey). The colored vertical lines show the different components of each line, with their respective number labeled, used to ascertain the component structure for photoionization modeling. The {\MgII} profile has 26 components out of which, component structure in the central saturated region is ascertained from the unsaturated {\CaII} doublet lines, the {\FeII} multiplets, and the {\MgI} line. The {\FeII}$\lambda2586$~{\AA} and {\FeII}$\lambda2374$~{\AA} lines fall in the {\Lya} forest region. }
    \label{system_plot}
   \end{figure*}

For the analysis of the absorption at $z_{abs} = 1.13331$, we consider only those ions with at least one transition falling at wavelengths outside of the {\Lya} forest. We refer to these features as \textit{secure lines} for which estimates or limits on column densities unaffected by contamination are possible. The integrated column densities of an entire absorption feature for all the detected metal absorption lines are estimated using the apparent optical depth (AOD) method of \citet{1991ApJ...379..245S} and given in Table~\ref{tab:A2}. The system plot of Figure~\ref{system_plot} centered on the redshift of the host galaxy ($z_{gal} = 1.1334$) shows these lines. 
The {\MgIIdblt}~{\AA}, {\MgI}~$\lambda 2852$~{\AA}, {\CaII}~$\lambda\lambda3934, 3969$~{\AA}, and {\TiII}~$\lambda\lambda3242, 3384$~{\AA} absorption features fall outside of the {\Lya} forest and are detected at $\geq 5\sigma$ significance. Absorption from {\FeII} in nine different transitions from $2600$~{\AA} to $1901$~{\AA} are all in the forest region ($\lambda \lesssim 5750$~{\AA}) and are contaminated to different levels by the IGM over the $\Delta v \sim 460$~{\kms} range that the {\MgII} absorption spans. 


The rest-frame equivalent width $W_r (\MgII~\lambda2796$) = 3.185~${\pm}$~0.032~{\AA}, and velocity spread of $\Delta v(\MgII~\lambda2796$) $\sim 460$~{\kms} along the line of sight makes this an ``ultra-strong {\MgII} absorber'', following the nomenclature scheme of \citet{2007ApJ...658..185N}\footnote{$\MgII$ absorbers with rest-frame equivalent width of the 2796~{\AA} transition in the range of 2.7~{\AA} $\lesssim W_r \lesssim 6.0$~{\AA}}. The absorption in {\MgII} is made up of several kinematic sub-components. The profile fitting (see \S\,\ref{ion_model}) decomposes the absorption into 26 distinct kinematic components. The respective parameters with the component velocities are listed in Table \ref{tab:abs-properties} (in the appendix), and the corresponding Voigt profile model is shown in Figure \ref{system_plot}. The Voigt profile fit was obtained using the \textsc{VoigtFit} package v3.21.7 \citep{2018ascl.soft11016K}, by applying Voigt-Hjerting fits to the continuum-normalized spectrum of the quasar. Initially, the velocity structure across available transitions was examined to identify the components (using least saturated transitions like {\MgI} and {\CaII} for heavily blended regions), assuming a consistent component structure and starting with a minimum number of components, the fits are iteratively defined and a final component structure explaining the absorption profile with the least amount of complexity is finalized. The {\MgIIdblt~{\AA}} absorption in the [$-$$30$, +120]~{\kms} velocity range shows saturation. The kinematic sub-components in this range (components 12 to 21) were constrained based on the component structure of the unsaturated {\MgI}~$\lambda2852$~{\AA}, and {\CaII}~$\lambda\lambda3934, 3969$~{\AA} lines, with the velocity of the components allowed to vary during the fitting process. 


We also estimate the kinematic spread $\omega_v$\footnote{Kinematic spread is defined as the second velocity moment of the pixel optical depth-integrated over the velocity range of an absorption feature, i.e., $\omega_v^2 = \int_{-v}^{+v} v^2 \tau_a(v) dv / \int_{-v}^{+v} \tau_a(v) dv$, where $\tau_a(v) = \ln (f_c(v)/f(v))$ is the apparent optical depth of each pixel of an absorption feature within the interval [-v, +v]~{\kms}, where $v = 0$~{\kms} corresponds to the redshift of the absorber.}, of the {\MgII} and {\MgI} absorption. Compared to the equivalent width which is biased towards the strength of individual components in an absorption feature, $\omega_v$ is a more sensitive measure of the velocity spread, since it scales as $v^2$. We estimate $\omega_v (\MgII~\lambda2796~{\AA}) = 110~\pm~9$~{\kms}. The kinematic spread of a strong absorber population indicates how the absorption components are spread out in velocity space, and by extension, the relative opacity of the individual component absorbers and is found to have two kinds of distribution in equivalent width. Low values of kinematic spread for a given equivalent width would result when the profiles are opaque over a significant portion of the absorption velocity range, with very few high-velocity unsaturated sub-components. On the other hand, high values of kinematic spread result from complex profiles with several weak components separated in velocity. \citet{2007ApJ...669..135M} suggested that, at low-$z$, the latter would be the case when these strong {\MgII} absorbers are tracing disjoint parcels/filaments that form a stream such as in tidal debris \citep[e.g.,][]{D_Onghia_2016}. In contrast, continuous saturated absorption of low ionization gas over a large velocity range can result from the line of sight piercing through an unbroken stream of gas, such as the cool phase entrenched within a galactic wind either in its outflow or later re-accretion stage \citep{2001ApJ...562..641B,2011MNRAS.412.1559N}. The absorber's {\MgIIdblt}~{\AA} kinematic spread and profile structure are consistent with this latter scenario.

The {\MgI}~$\lambda2852$~{\AA} absorption is detected in 14 of the 26 velocity components seen in {\MgII}, with a kinematic spread of $\omega_v = 61~\pm~5$~{\kms}. The {\CaII}~$\lambda\lambda3934, 3969$~{\AA}, and {\TiII}~$\lambda3384$~{\AA} are detected at $\geq 5\sigma$ at those central velocities where the {\MgII} absorption is strongest (see Figure \ref{system_plot}). These lines are weaker when compared to the lines from {\MgI} and {\MgII} because of the low cosmic abundance of the respective elements, their possible depletion onto dust \citep{1990ARA&A..28...37M, 2016A&A...591A.137G}, and also because of the ionization energies of {\CaII}, and {\TiII} being lower than {\HI} (11.9 eV, and 5.9 eV, respectively) which leads to their low ionization fractions in photo-ionized diffuse gas. Meanwhile, their detections also indicate that the line of sight is probing a high {\HI} column density environment such as a Damped {\Lya} Absorber (DLA, $\log [N(\HI)/\cmsq] \geq 20.3$), or a sub-DLA ($19.0 \leq \log [N(\HI)/\cmsq] < 20.3$). The {\HI} Lyman series lines are all below the wavelength coverage of UVES, and hence we do not have any direct information on the hydrogen associated with this system. For estimating metallicity, we have used the {\HI} column density generated via component-by-component modeling, as explained in section~\ref{ion_model}.

The detection of {\CaII} with $W_r \approx 223$~m{\AA}provides some useful pointers on the physical origin and the {\HI} content of this absorber. Observations of intermediate and high-velocity gas in the halo of the Milk Way find a covering fraction (area filling factor) of $\sim 40 - 60\%$ for {\CaII}, with physical properties consistent with that of strong {\MgII} absorbers, suggesting the two ions as co-existing, tracing the same phase of the gas in the CGM of galaxies \citep{Ben_Bekhti_2012,Bish_2019}. In a study of 23 {\CaII} absorbers in the redshift interval 0.004 $< z <$ 0.474, \citet{2011A&A...528A..12R} found that absorbers with $W_r(\CaII~\lambda3934$\AA) $>$ 35~m{\AA} ($\log (N/\cmsq) > 11.5$) are consistent with tracing optically thick gas with $\log~(N(\HI)/\cmsq) > 17.2$, with stronger systems ($W_r \gtrsim 200$~m{\AA}) related to DLAs \citep{2007MNRAS.379.1409Z}. Furthermore, they found that {\CaII} shows an increasing effect of depletion due to dust with increasing {\HI} column density. Other high resolution spectroscopic studies also \citet{Robertson1988} in the first census of extragalactic {\CaII} absorption, found that the {\CaII} is severely underabundant compared to both {\FeII} and {\MgII}, typically smaller by a factor of 100. They argued that the underabundance suggests a higher tendency for the {\CaII} to condense onto dust grains compared to {\MgII} and {\FeII}.

The UVES spectrum covers several prominent {\FeII} lines from $1901$~{\AA} to $2600$~{\AA}. However, all these lines are contaminated to varying levels by the  {\Lya} forest. The extent of contamination can be discerned from the comparison of the apparent column density profiles amongst these lines, and with the {\MgII}. It can be seen from Figure\,\ref{pie_model} that the {\FeII}~$\lambda2344$~{\AA} line closely follows the {\MgII} across 21 out of the 26 components spread over the velocity range of the absorption, implying little contamination from the {\Lya} forest. We therefore use this $2344$~{\AA} line to model the component structure in {\FeII}. The column density ratio of {\FeII} to {\MgII} across the components can serve as a strong constraint on the density of the gas (e.g., \citealt{2002ApJ...565..743R}). It also helps in determining the chemical enrichment history, since the [Mg/Fe] abundance is linked to the overall rates of core-collapse and Type Ia SNe respectively (e.g., \citealt{2015MNRAS.451.1806D}). In the ionization models discussed in Sec\,\ref{ion_model}, {\FeII}~$\lambda2344$~{\AA} was also used in constraining the model predicted profiles when it is determined to be unblended based on similarity with {\MgII}, and the strengths of other observed {\FeII} transitions. 

  \subsection{Ionization Modelling}\label{ion_model}

We use the spectral synthesis routine {\CLOUDY} v23.01 \citep{Gunasekera2023} to constrain the ionization properties of the absorber. The individual components in the {\MgII} absorption are considered separately in the models. The models are generated assuming each absorbing component to be an isothermal plane-parallel slab of constant density, with little dust content, and following a solar abundance pattern~\citep{grevesse2011chemical}. We assume little dust content to simplify the parameter space. Since Ca is known to be affected by depletion onto dust, the {\CaII} was not included as species constraining the ionization models. The absorber is assumed to be photoionized by the extragalactic UV background radiation (UVB) for the absorber redshift of $z = 1.13$. The UVB is based on the ionizing background model of \citet{2019MNRAS.484.4174K} (hereafter KS19), which is an improvement over the \citet{2012ApJ...746..125H} model, incorporating current estimates of {\HI} distribution in the IGM \citep{2014MNRAS.442.1805I}, recent estimates of QSO emissivities, cosmic star formation rate density, and far-UV extinction from dust \citep{2015MNRAS.451L..30K}. 

For ionization modeling, we use the component-by-component Bayesian modeling (CMBM) approaches described in \citet{Sameer2021,Sameer2022}. Briefly, a grid of {\CLOUDY} photoionization thermal equilibrium models is generated with metallicity, hydrogen number density, and neutral hydrogen column density varying over the range of {\metallicity} $\in$ [$-$4.0, 2.0], {\hdenu} $\in$ [$-$6.0, 2.0], and {\coldenu} $\in$ [11.0, 23.0] with step-size of 0.1~dex. The neutral hydrogen column density sets the stopping criterion for each of the converged models. An initial VP fit model to the unsaturated components observed in \MgI, {\MgII} and/or {\FeII} ions determines the component structure i.e. the absorption centroids and the total Doppler broadening parameter - which comprises of thermal and non-thermal broadening. These ions are referred to as ``constraining'' ions. Each cloud in the absorption system is parameterized using \metallicity, \hden, \colden, absorber redshift $z$, and non-thermal broadening parameter $b_{nt}$. The converged model also predicts the photoionization equilibrium temperature. Using this temperature, and the $b$-measured for the constraining ion during the initial Voigt profile fitting, the non-thermal contribution to line broadening is determined.   Each component in the absorption system is characterized by five parameters: $\log Z$, $\log n_{\rm H}$, $\log N(\rm HI)$, $z$, and $b_{nt}$. We use a uniform prior on $b_{nt}$ (the non-thermal contribution to the Doppler parameter of a cloud) ranging between [0, $b+3\sigma(b)$] where $b$ is the Doppler broadening parameter determined from a preliminary Voigt profile fit. We determine the Doppler broadening parameter $b$ for all transitions using the equation $b^{2}=b^{2}_{\mathrm{nt}}+b^{2}_{\mathrm{t}}$, where $b_{\mathrm{t}}=\sqrt{2kT/m}$ is the line broadening due to temperature and $b_{\mathrm{nt}}$ the line broadening due to non-thermal effects. The non-thermal broadening component is assumed to be the same for all the transitions in the same cloud.

This information is coupled with the column densities of various ions in {\CLOUDY} models to generate synthetic absorption profiles convolving with the instrumental LSF. These models are sampled from the interpolated {\CLOUDY} grid, using PyMultinest\footnote{\url{https://github.com/JohannesBuchner/PyMultiNest}}\citep{2014A&A...564A.125B}, to determine the parameters where the loss function (difference between data and model formulated as a chi-squared) is minimized in all the observed transitions. The posterior distributions for the physical parameters of the model are determined by adopting uniform priors on \metallicity, \hden, and {\colden} for the range of values covered in the {\CLOUDY} grid.

\smallskip

\begin{figure*}
    \centering
    \includegraphics[width=1.0\textwidth]{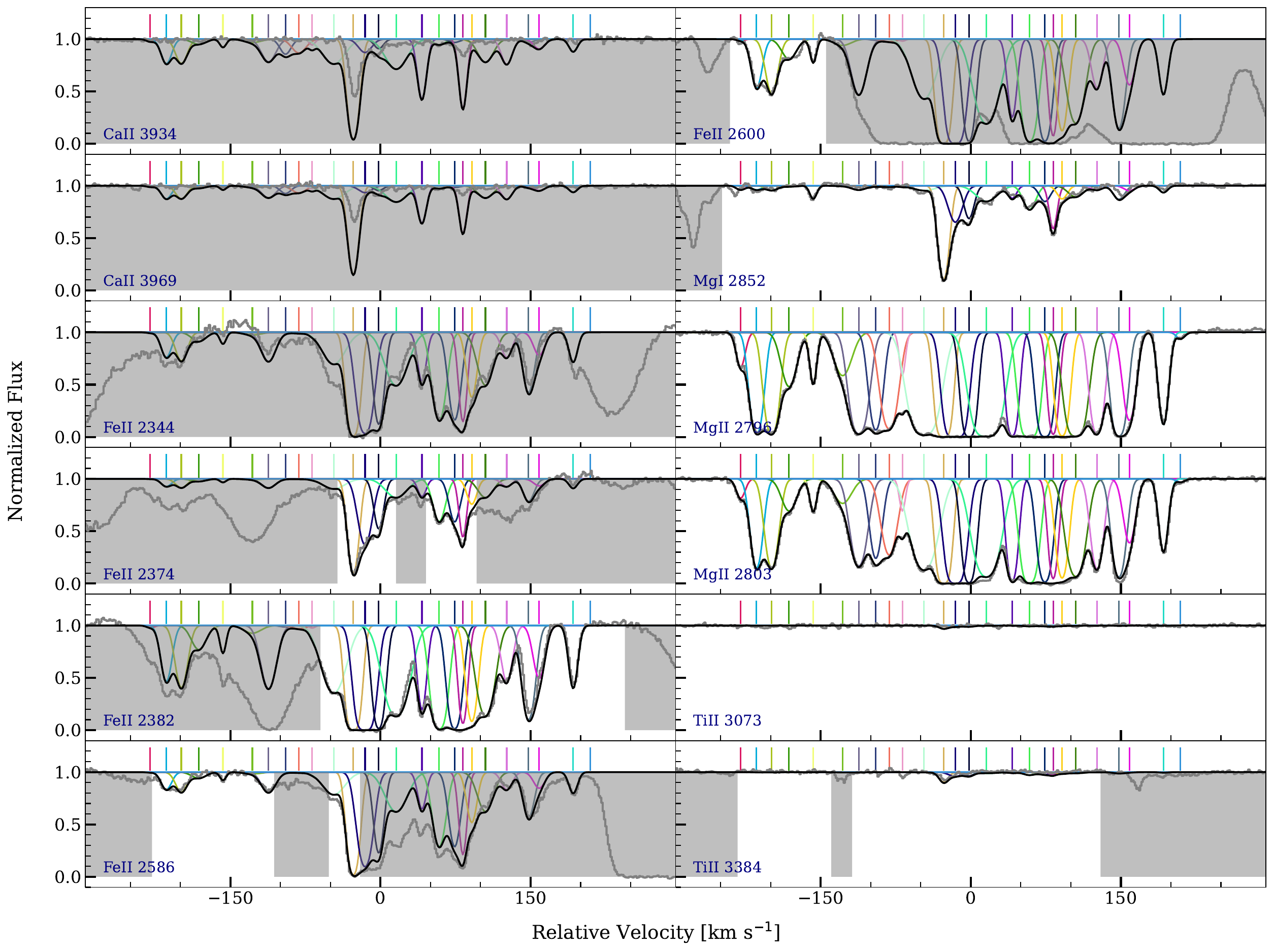}
    \caption{This figure shows the component profiles of metal lines predicted by photoionization models for the 26 components seen in {\MgII}. The modeling results shown here are summarised in Table~\ref{tab:cloudproperties}.
    }
    \label{pie_model}       
\end{figure*}

The component-by-component model results are given in Table \ref{tab:cloudproperties}. In Figure \ref{pie_model}, the synthetic profiles of the individual components generated for the converged models are shown. The hydrogen Lyman series lines are not covered by the spectrum. In the absence of {\HI} information, it is not possible to constrain the metallicity of the clouds directly. Instead, the photoionization models were employed to make predictions regarding the {\HI}. The inferred posterior distributions of parameters for the different clouds are visualized in the violin plots presented in Figure~\ref{pie_model} (e.g., \citealt{2022MNRAS.514.6074N, Sameer2024a}). 


\smallskip

The densities in the individual components are constrained by photoionization models to be in the range $n(\H) \approx 10^{-3} - 2.5$~{\cc}, with higher densities associated with components that are also detected in {\MgI} and {\CaII}. Metallicity is predicted to span a large range, with several of the components having supersolar values (see the panel with the posterior distribution for metallicity in Figure \ref{pie_model}). The uncertainty in the {\FeII} column density due to contamination from the {\Lya} forest would be the biggest source of uncertainty in these metallicity predictions, especially for those components for which there is no {\CaII}. Hence, we treat these metallicity predictions with caution, in absence of the {\hi} coverage. We note that our results assume a solar abundance pattern; relaxing this assumption will add further complexity to our model and is beyond the scope of this work. The total uncertainties reported in Table~\ref{tab:cloudproperties} include only statistical uncertainties. The model predicted {\HI} column densities for the individual components when added together yields \colden $\approx 22.5$. The total {\HI} column density given by the model is consistent with the more general trend of ultra-strong {\MgII} absorbers being associated with sub-DLAs, and DLAs. For \colden $\gtrsim 19.5$, the ionized hydrogen fraction is negligible and therefore the integrated {\HI} column density is approximately equal to the baryonic column density \citep{Jenkins_2009,2015ApJ...804...83S}. In addition to the predicted {\colden}, in Table~\ref{tab:abs-properties}, we report the model predicted column densities of ions detected at the 3$\sigma$ level. We find that the model predicted total column densities are consistent within one sigma for {\MgI} and {\TiII}, and with the lower limits for {\MgII} and {\FeII} (see Table \ref{tab:A2}). {\CaII}, on the other hand, shows depletion by a factor of $\sim$6.

The models also show the absorption to be a composite of several components with an interquartile range of 10 pc - 1 kpc thickness. Consequently, it appears that the line of sight is tracing a kinematically clumpy medium, with metals concentrated into zones of the scale of parsecs, rather than uniformly dispersed along a single coherent absorbing structure. The discrete nature of low ionization clouds in the CGM is in line with the more direct measurements of absorber sizes, and metallicity difference between components through gravitational arc tomography and lensed quasar observations of individual halos at closely spaced angular scales. \citep{2001ApJ...562...76R,2004ApJ...615..118E,2018Natur.554..493L,2018ApJ...853...95R,2019ApJ...886...83K,2021MNRAS.505.6195A}, and also results from CGM simulations where small-scale over-dense regions are traced by {\MgII} \citep{2021ApJ...923...56D}.

\section{Emission line analysis of MUSE data}\label{sec:sec4}

The $1^{\prime} \times 1^{\prime}$ MUSE FoV, corresponds to $500$~kpc~$\times500$~kpc of projected separation at the redshift of the absorber.  In the blind emitters catalog of \citet[see section~3.1]{2021MNRAS.508.5612M}, only one galaxy is identified within $\pm 5000$~{\kms} of the {\MgII} absorber redshift. The galaxy is at a projected separation of $\rho \approx 18$~kpc from the quasar line of sight. The top-right panel of Figure \ref{white_light} shows the MUSE pseudo-narrow band image of the galaxy obtained by cropping the continuum subtracted sub-cube over a $20$~{\AA} wavelength range centered on the redshifted position of the [\OII]~$\lambda \lambda~3727,~3729$~{\AA} emission. As explained earlier, the quasar's PSF in the MUSE field was subtracted to remove the quasar light contaminating the galaxy emission. The galaxy orientation and kinematics are modeled using the 3D line fitting package GalPaK$^{3D}$ \citep{2015AJ....150...92B}. GalPak$^{3D}$ generates 2D velocity and flux maps for the galaxy and provides constraints on the inclination of the galaxy with respect to the plane of the sky ($i$), the position angle of the galaxy major axis (PA)\footnote{Position angle is the angle that the major axis of the galaxy makes with the North Celestial Pole, measured along the direction of RA.}, the maximum circular velocity of the emitting gas ($v_{max}$\footnote{This is the max/peak velocity determined from the inclination corrected rotational velocity profile of the gas assuming circular rotation}), the velocity dispersion ($\sigma_v$), and the half-light radius ($R_{1/2}$).

Figure ~\ref{vmap} shows the inclination corrected velocity map for the galaxy and the relative position of the quasar line of sight. The position coordinates of the galaxy, and the morpho-kinematic parameters obtained from GalPak$^{3D}$ (highlighted in cyan) modeling are tabulated in Table~\ref{tab:3}. These values were further used to estimate the dynamical mass (M$_{dyn}$), the halo mass M$_h$, and stellar mass M$_*$. 

\begin{table}[hbtp]
\caption{Properties of the Galaxy Associated with ultrastrong \MgII}
\begin{center}
            \begin{tabular}{l r}
            \hline
            \hline
            Property & Value \\
            \hline
            z & 1.1334\\
            RA & 16:21:16.7\\
            Dec  & $-00:42:52.8$\\
            \rowcolor{LightCyan}Flux$_{[\OII]}$($10^{-17}~\mathrm{erg~s^{-1}~{\cmsq}}$) & $6.0\pm0.1$\\
            \rowcolor{LightCyan}$R_{\frac{1}{2}}$(kpc) & $5.7\pm0.1$\\
            \rowcolor{LightCyan}Inclination & $54^{\circ}\pm1$\\
            \rowcolor{LightCyan}P.A. & $32^{\circ}\pm1$\\
            \rowcolor{LightCyan}V$_{max}$(\kms) & $186\pm11$ \\
            \rowcolor{LightCyan}r$_{t}$(kpc)$^a$ & $5.4\pm0.6$\\
            \rowcolor{LightCyan}$\sigma$(\kms) & $42\pm2$\\
            $\alpha$ & $21.4^{\circ}\pm0.9$\\
            M$_{dyn}$( $10^{10}$ M$_{\odot}$ ) & $4.6\pm0.5$\\
            M$_h$( $10^{12}$ M$_{\odot}$ ) & $1.1\pm0.2$\\
            M$_*$( $10^{10}$ M$_{\odot}$ ) & $4.700\pm0.002$\\
            SFR(M$\odot$ yr$^{-1}$) & $8.3~\pm~2.1$\\
            sSFR(Gyr$^{-1}$) & $0.18~\pm~0.04$\\
            R$_{vir}$(kpc) & $152.3\pm9.0$\\
            \hline
            \end{tabular}
\end{center}
\label{tab:3}
\begin{flushleft} \small
Notes-- A seeing of 0.62$^{\prime\prime}$ was estimated using PampelMuse. The galaxy parameters are determined from the 3D fits using the GalPaK$^{3D}$ algorithm. The azimuthal angle ($\alpha$), the galaxy stellar, halo, and dynamical masses, the SFR, and the specific SFR were estimated from the galaxy parameters.\\
$^a$ Turnover radius for the velocity rotation curve
\end{flushleft}        	
\end{table}

The mean wavelength of the [\OII]~$\lambda\lambda3727,~3729$~{\AA} emission yields a galaxy redshift of $z_{gal} = 1.1334~{\pm}~0.0001$. Fitting the [\OII] doublet emission with a Gaussian double peak model also yields a similar redshift. The [\SII]~$\lambda\lambda4068,~4076$~{\AA} emission lines are non-detections. Other commonly seen emission lines such as H$\beta$, H$\gamma$, and [\OIII]~$\lambda4363$~{\AA} are outside of the MUSE wavelength coverage. The star-formation rate (SFR) given in Table ~\ref{tab:3} was determined using the empirical [\OII] - SFR calibration relation given by equation 4 of \citet{2004AJ....127.2002K}, using the intrinsic Luminosity determined from equation 18 of the same article.. The extinction corrected star formation rate thus estimated places this galaxy slightly above but within the $\approx 0.3$~dex scatter about the average for the population of star-forming galaxies represented by the galaxy main-sequence at $z \sim 1$ (e.g., \citealt{2015A&A...577A.112L}). Using the galaxy $R_{\frac{1}{2}}$ size and maximum rotation velocity, the galaxy's dynamical mass within its half-light radius and its halo mass are estimated (see Table \ref{tab:3}). For the Halo Mass estimation, equation 1 of \citet{Bouch__2016} was used and the resultant halo mass was used to determine the stellar mass via a Stellar-Halo mass scaling relation(equation 6 of \citet{Girelli_2020} derived from \citet{2010ApJ...710..903M}). The dynamical mass was also determined from \citet{Bouch__2016} described in section 4.5. The azimuthal angle of $\alpha \approx 21^{\circ}$ indicates that the line of sight is probing a region close to the kinematic major axis of the galaxy (see Figure \ref{vmap}).

\section{Origin of the Ultra-strong MgII Absorption}\label{origin}

Strong {\MgII} absorbers are found to trace both isolated galaxies and dense group/cluster environments  \citep{Nestor_2007,2011MNRAS.412.1559N,2018MNRAS.474..254R,2021MNRAS.503.4309L, Gauthier_2013,2021MNRAS.502.4743H,10.1093/mnras/stz3590}. When multiple galaxies are present, tidal interactions and the superposition of halos along the line of sight are expected to give rise to kinematically complex and strong absorption \citep{2011ApJ...743...10B,2018ApJ...869..153N,2020MNRAS.499.5022D,10.1093/mnras/stac1824}. At the same time, the covering fraction of strong {\MgII} is found to be significant ($\approx 60 - 80$\%) even for galaxies without a companion out to $\rho \approx 100$~kpc, as long as the impact parameter of the galaxy with the line of sight is not very large ($\rho \lesssim 40$~kpc, \citealt{2019MNRAS.490.4368S}).  In a study of 20 ultrastrong {\MgII} absorbers and their host galaxies at $z \approx 0.5$,  \citet{2022MNRAS.513.3836G} find that as much as one-third of the ultrastrong {\MgII} absorber population are linked to isolated galaxies. Such results imply that, alongside the environmental factors, intrinsic properties of a galaxy such as its star formation rate \citep[e.g.,][]{2021MNRAS.502.4743H}, luminosity \citep[e.g.,][]{2010ApJ...724L.176C}, and processes such as inflows and outflows that define the baryonic cycle can also enhance the physical cross-section of the {\MgII} absorbing halos. For the absorber discussed in this paper, we do not find any evidence for the line of sight tracing a galaxy overdensity environment at the redshift of the absorber. The one galaxy identified by MUSEQuBES appears to be isolated, with no bright companion within a $\rho \approx 500$~kpc of projected separation covered by the MUSE field-of-view. The velocity map of the galaxy has the pattern of a differentially rotating disk (see Figure \ref{vmap}) ruling out close interactions with neighboring galaxies, if any. We now discuss some plausible scenarios for the origin of the {\MgII} absorber based on the comparison between absorber and galaxy properties. 

\subsection{Can the Absorption be Tracing Active Outflows?}

Star-formation-driven outflows and cold gas accretion are major contributors to the strong {\MgII} absorber population (e.gs., \citealt{2006MNRAS.371..495B}; \citealt{2009ApJ...703.1394M}; \citealt{2010MNRAS.403..906N}; \citealt{2016ApJ...833...39S}, \citealt{2019MNRAS.490.4368S}). This is observed through a bimodality in the distribution of {\MgII} absorbers in relation to the azimuthal angle \citep{2012MNRAS.426..801B,2012ApJ...760L...7K,2019MNRAS.485.1961Z}. A higher incidence of {\MgII} absorbers is found for galaxies at azimuthal angles $55^{\circ} < \alpha < 90^{\circ}$ and inclination i $<$ 60$^\circ$, where the line of sight is probing regions closer to the projected minor axis of the galaxy along which bi-conical outflows are channeled. Similarly, along $0^\circ < \alpha < 35^\circ$ corresponding to the projected major axis, a higher incidence of {\MgII} is found because of the line of sight intercepting an extended disk formed from accreting intergalactic gas or material falling back from past outflows.

\begin{figure*}[hbp]
    \centering
    \includegraphics[width=1\textwidth]{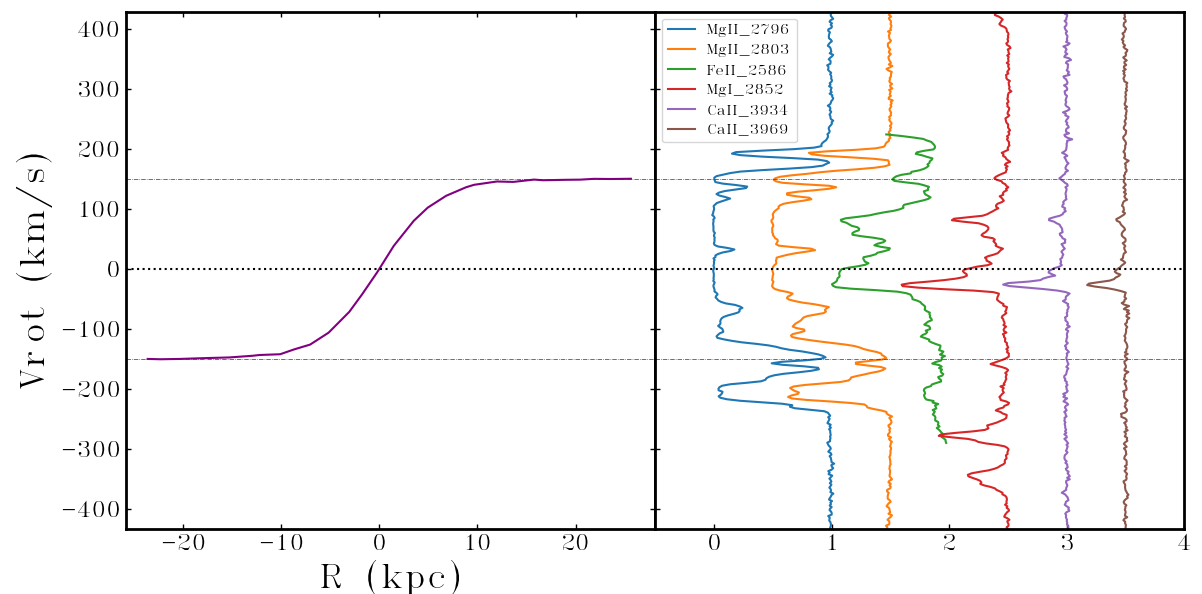}
    \caption{{\tt Left:} The inclination-angle-corrected rotation curve obtained from GalPaK$^{3D}$. {\tt Right:} The detected absorption profiles with normalized flux (+ offset) along the X-axis. We exclude all FeII features except the $\lambda2586$ {\AA} line, as the others are too saturated or have low S/N.This line is also truncated on both sides in the plot due to contamination from the forest, making it visually cleaner for comparison.   The dotted lines mark the range of velocities covered by the rotation curve. The central dotted line at 0 {\kms} is for the redshift of the galaxy.}
    \label{vrot}
\end{figure*}

\begin{figure*}[hbp]
    \centering
    \includegraphics[width=0.4\linewidth]{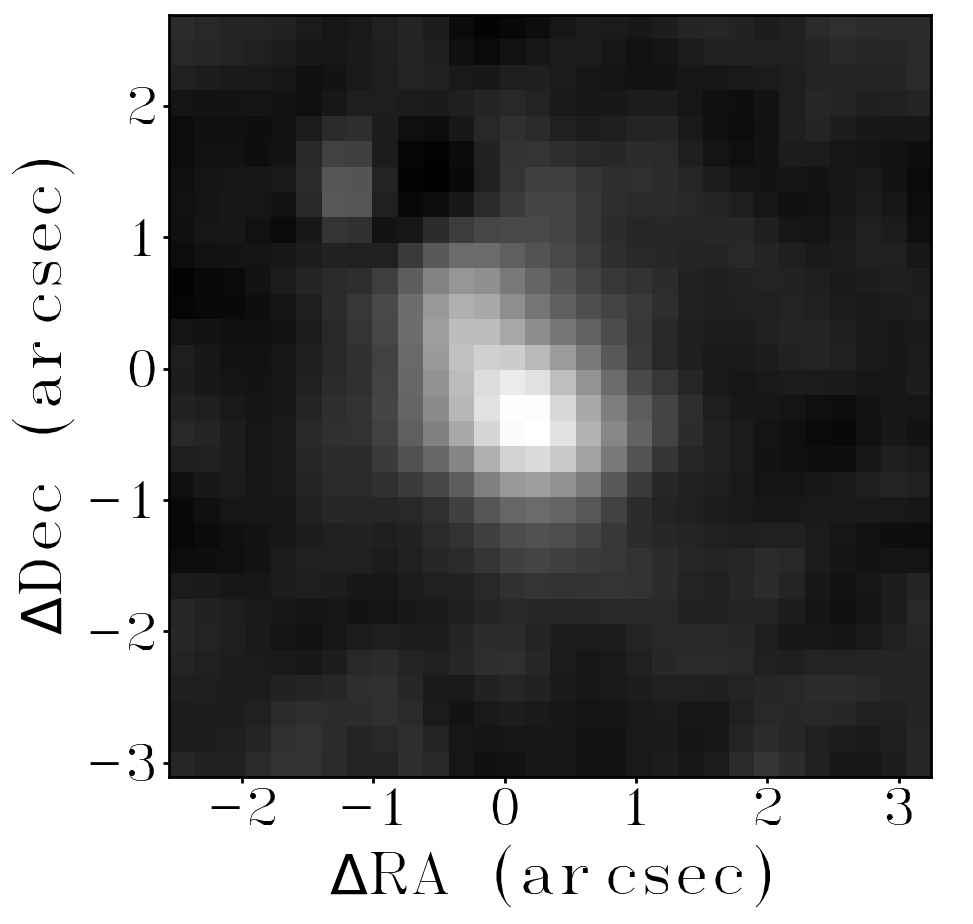}
    \includegraphics[width=0.5\linewidth]{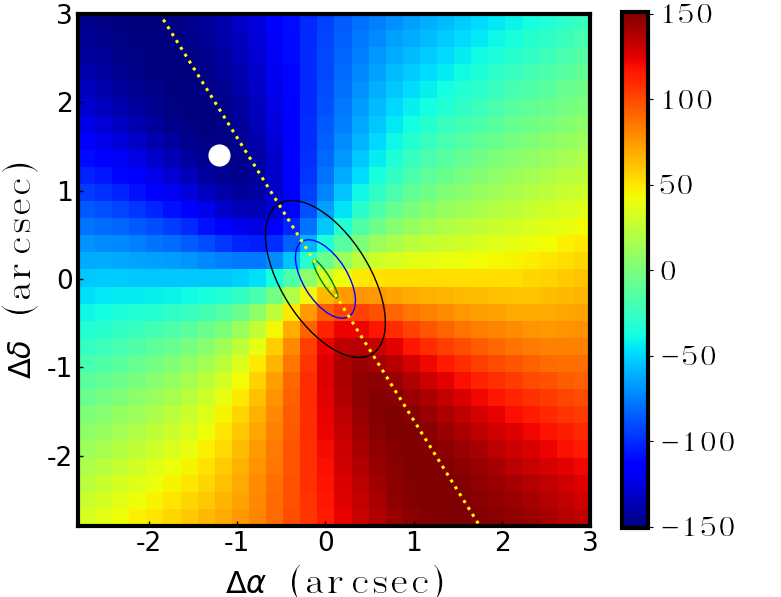}
    \caption{\textit{Left}: MUSE pseudo-narrow band image spanning the wavelength range $7940 - 7960$~{\AA} centered on the redshifted position of the [\OII]~$\lambda\lambda3727, 3729$~{\AA} emission. The faint residuals after removing the quasar light are seen at the \textit{top-left} corner of the image.A Gaussian filter of 2 sigma has also been applied to smoothen the image. \textit{Right}: The galaxy velocity field map derived using Galpak$^{3D}$ along with the quasar line of sight (\textit{white filled circle}). The color grid indicates the corresponding velocities in the map. The \textit{yellow dotted line} represents the major axis of the galaxy at the position angle. The concentric ellipsoids define the iso-flux contours. The azimuthal angle between the quasar line of sight and the galaxy's major axis is $\alpha \approx 21.4^{\circ}$. The quasar line of sight probes a region close to the projected major axis where the emitting gas has a velocity of $\approx -150$~{\kms}, tracing the flat part of the rotation curve.}
    \label{vmap}
    \end{figure*}

In the case of the absorber discussed here, the azimuthal angle of $\alpha = 21^{\circ}$ is further away than what is required to plausibly intercept an outflow. The empirical expression in \citet{Boogaard_2018} relating SFR with $M_*$ places the absorber host galaxy along the main sequence of star-forming galaxies where the SFR is below the threshold of a starburst required for expelling large quantities of gas.
Using the half-light radius and SFR, we estimate the star formation rate surface density\footnote{$\Sigma_{\mathrm{SFR}}$ is defined as $\mathrm{SFR}/(2\pi R_*^2)$, where $R_*$ is the half-light radius.}  to be $\Sigma_{\mathrm{SFR}} = 0.041$ ~M$_\odot~\mathrm{yr^{-1}~kpc^{-2}}$, which is also below the characteristic value of $0.1$~M$_\odot \mathrm{yr^{-1}~kpc^{-2}}$ required for large-scale galactic winds \citep{Heckman_2015, 2011hst..prop12603H}. This however does not eliminate the possibility of the absorber tracing gas deposited into the CGM from past outflow events when the SFRs were plausibly higher. In fact, some fraction of the {samples in studies of ultrastrong {\MgII} systems are associated with post-starburst galaxies (e.g., \citealt{2011MNRAS.412.1559N}, \citealt{2022MNRAS.513.3836G}). For the absorber presented here, the velocity of V$_{\mathrm{gas}} = \sim 228$~{\kms} with which the absorbing gas is moving does not exceed the halo escape velocity of V$_{\mathrm{esc}} \approx 466$~{\kms} of the galaxy to the extent of becoming gravitationally unbound. Such expelled gas is likely to be retained as halo gas itself since the infall timescales can be of the order of or larger than Hubble time \citep{2020MNRAS.497.4495M}. In the above-mentioned calculation of escape velocity (Equation 7 of \citealt{2016ApJ...833...39S}, section 4.6 of \citealt{2005ARA&A..43..769V}), V$_{\mathrm{gas}}$ is the velocity spread of the {\MgII} absorption relative to the galaxy rest frame assumed to be the outflow velocity V$_{\mathrm{out}}$. The halo escape velocity V$_{\mathrm{esc}}$ is determined from the maximum rotation velocity of the galaxy, and its virial radius. These quantities were used to estimate the gas escape fraction V$_{\mathrm{out}}$/V$_{\mathrm{esc}}$, a crucial factor in determining whether outflowing gas can escape the galaxy's gravitational well. For this case, the ratio of V$_{\mathrm{out}}$/V$_{\mathrm{esc}} \simeq 0.49$ suggests that the observed gas is unlikely to be escaping the galaxy but rather is falling back onto it.


\subsection{Absorption and Emission Kinematics}\label{kinematics}

The location of the absorption given by the azimuthal angle suggests that the absorption is nearly co-planar with the galaxy. We now compare the absorber and galaxy kinematics to understand whether the gas is part of CGM that is co-rotating with the disc. Figure~\ref{vrot} shows the absorption profiles of the key ions against the inclination corrected rotation curve for the galaxy given by GalPaK$^\mathrm{3D}$. The {\MgII} absorption span a velocity range of $\approx [-245, 225]$~{\kms}}, which is wider than the differential rotation of the disk. The velocity field map of the galaxy of Figure ~\ref{vmap} shows the quasar line of sight intercepting a region with approaching velocities of $\approx -150$~{\kms} tracing the flat parts of the rotation curve. Though the strongest absorption components in {\MgI}, {\CaII}, and {\FeII} share the same velocity as the emitting gas, the full range of {\MgII} absorption lies on both sides of the galaxy's systemic redshift indicating that the absorbing gas does not entirely follow the kinematics expected for disk rotation. 

\begin{figure}[htbp]
    \centering
    \includegraphics[width=1\linewidth]{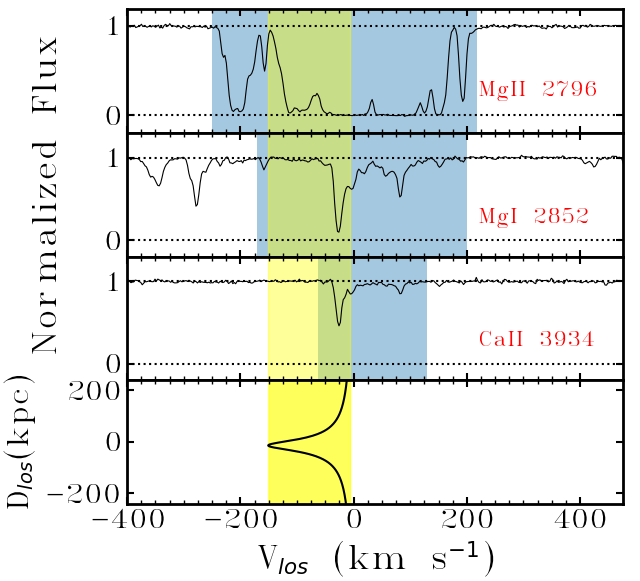}
    \caption{The bottom panel is the projected rotation velocity of the model galaxy V$_{los}$ along the line of sight D$_{los}$, based on the co-rotating disc model given in \citet{Steidel_2002}. The model parameters are described in \ref{kinematics}. The top three panels are the primary absorption lines with the blue box highlighting their absorption width and the yellow box highlighting the velocity span covered by the co-rotating disk Model. The 0 \kms is at the systemic galactic redshift. As can be seen, only a narrow range of velocities can be explained by a co-rotating disk-halo structure}
    \label{steidel}
\end{figure}

To investigate this further, we use the kinematic thick disk model of \citet{Steidel_2002} and compare the absorber gas kinematics with typical rotating disks to see if the entire velocity range of the absorption can be attributed solely to the disk and a co-rotating inner halo. For a gas disk thickness of $H_\mathrm{eff}$, the model provides the projected circular velocity along the line of sight at a given impact parameter along the galaxy's major axis. Such a projected rotation curve along the line of sight for model galaxy parameters of $H_\mathrm{eff} = 100$~kpc, and $h_v = 1000$ \footnote{The corotating thick-disk model has two free parameters-$H_\mathrm{eff}$ is the effective thickness in the z-direction and $h_v$ is the velocity scale height/factor adjusted to define the exponentially declining velocity parameter $v_{\phi}=v_c e^{-(|z|/h_v)}\hat{\phi}$ where $v_{\phi}$ is the tangential velocity }is shown in Figure \ref{steidel} bottom panel. It is found that even for such very large disk thicknesses, it is difficult to reproduce the full velocity range of the {\MgII} absorption. However, a portion of the strongly saturated region of {\MgII}, which also coincides with the strongest components in {\MgI}, {\CaII}, and {\FeII}, is consistent with the disk rotation. Since much of the absorption profile is not encompassed within the rotation curve, it has to be concluded that extraplanar gas with random velocities might also be contributing to the observed wide velocity spread. Such a conclusion is consistent with both simulations and observations that find such additional components corresponding to gas flows and tidal streams as necessary to explain the full range of strong {\MgII} kinematics \citep{2017gefb.confE..20H,2020ApJ...904...76H,2010ApJ...711..533K,2016ApJ...824...24D}.  



\subsection{Absorption Tracing Halo Gas Structures}
Star formation events of the past can populate the inner halos ($R  < 0.5R_{vir}$) of galaxies with cold gas \citep[e.g.,][]{2015PhDT........94N}. Observationally one finds an increased covering fraction of metal-enriched gas with a large line of sight velocities ($\times$ hundred {\kms}) for galaxies having relatively high sSFR \citep[$\approx 10^{-7} - 10^{-8}$~yr$^{-1}$,][] {2023MNRAS.523.5624W}. This is attributable to the presence of more recycled gas in the CGM from past outflows. Simulations show that the recycling of such gas back into the ISM of galaxies happens at a lower efficiency in halos with $M_h \gtrsim 10^{12}~M_\odot$ with timescales for reaccretion comparable to or longer than the Hubble time \citep{2020MNRAS.497.4495M}. Consequently, much of the wind material persists within the halo itself.  Such recycled metals, especially in the denser and cooler phases of the CGM, can form a clumpy distribution resulting in metallicity and column density variations between components along a given line of sight (e.g., \citealt{2013ApJS..204...17W}, \citealt{2018MNRAS.481..835O}). In the photoionization models of the absorber discussed in this paper (Sec~\ref{ion_model}), the metallicities for several of the components are $\gtrsim Z_\odot$, consistent with gas that is enriched, with differences of as much as $\approx 0.5$~dex between components for metallicity, and $\approx 3$~dex for column density.

Using cosmological simulation data, \citet{2020ApJ...904...76H} show that the circumgalactic gas traced by {\MgII} can have a distribution symmetric about the rotation axis of the galaxy with an angular momentum aligned with the rotation axis. Such corotating gas can be spatially extended out to distances of $\gtrsim 0.5R_{vir}$ \citep{2019ApJ...878...84M,2020ApJ...904...76H}. With the quasar line of sight oriented along the galaxy's major axis and probing at small impact parameters, the ultra-strong {\MgII} absorption we observe most probably originates from a combination of such gas clouds within the halo and even beyond. When interpreting the broad velocity dispersion of the {\MgII}, it's important to remember that absorption components that are closely spaced in velocity need not necessarily coincide spatially~\citep[e.g.,][]{Peeples2019,Marra2024}. The {\MgII} doublets show continuous absorption across $\Delta v \approx 460$~{\kms}. Of this, absorption in the velocity range $-150~{\kms}\lesssim \Delta v (\MgII) \lesssim 0$~{\kms} (corresponding to $\approx 35$\% of the absorbing velocity pixels), shows kinematics consistent with gas co-rotating with the disk and extending up to a line of sight distances of $\approx 200$~kpc. The strongest absorption seen in {\MgI}, {\FeII}, and {\CaII} also trace this co-rotating gas. The remaining absorption portion must be tracing gas with velocities outside the co-rotating material.  Several parsec scale clouds situated even beyond $R_{vir}$ can contribute to absorption that is coincident in velocity with the inner CGM \citep{2015ApJ...802...10C}, depending on the velocity distribution. These two sources of absorption cannot be distinguished without more information. Some of the \MgII\ components with extreme positive and negative velocities w.r.t. the host galaxy can very well arise due to two-halo contributions from fainter galaxies that live in the same large-scale structure \citep[e.g.][]{2020ApJ...904...76H}.

\section{SUMMARY OF RESULTS}\label{sec:summary}

This study presents the analysis of an ultra-strong {\MgII} absorption associated with the circumgalactic medium of star-forming galaxy at $z_{gal} = 1.1334$ at an impact parameter of $\approx 18$ kpc. The key results are as follows: 

\renewcommand{\theenumi}{\roman{enumi}}%
\begin{enumerate}

   \item The {\MgII} absorption centered at $z = 1.13331$ has a rest-frame equivalent width of EW$_{0}^{2796}=3.185\pm0.032$~{\AA}, which makes it an ultrastrong {\MgII} absorber. The strong {\MgII} absorption feature is spread over $\Delta v \approx 460$~{\kms}. A Cloud-by-cloud ionization modeling decomposes this absorption in {\MgII} into 26 kinematically distinct components at the resolution of $VLT$/UVES. Coincident with the redshift of {\MgII} is also absorption from {\MgI}, {\CaII}, and {\FeII} detected at significance $\geq 5 \sigma$. The corresponding Lyman series lines are outside the wavelength coverage of UVES. 
   
   \item A component-by-component Bayesian photoionization modeling constrains the density and metallicity in the individual components to $n(\H) \approx 10^{-3}~-~2.5$~{\cc}, and  $-4.0 \lesssim \log (Z/Z_\odot) \lesssim 1.5$, with a large number of components having $\gtrsim Z_\odot$ (see Table~\ref{tab:cloudproperties}). The models predict an integrated {\HI} column density of $\log [N(\HI)/\cmsq] = 22.5$ for the absorber. 
   
   \item MUSEQuBES survey has identified only one galaxy within the $1^{\prime} \times 1^{\prime}$ MUSE FoV, and within $\pm 5000$ {\kms} of the absorber redshift. The absorber is at a projected separation of $\rho \sim 18$ kpc ($\rho/R_{vir} \sim 0.12$) from the galaxy, with $z_{gal} \approx z_{abs}$. The galaxy has a $M_* = 4.7 \times 10^{10}$~M$_\odot$, $M_h = 1.1 \times 10^{12}$~M$_\odot$, and an extinction corrected SFR = $8.3$~M$_\odot$~yr$^{-1}$, which places it slightly above the average for star-forming galaxies at $z \sim 1$. The azimuthal angle of $\alpha = 21.4^o$ suggests that the absorption is nearly co-planar with the galaxy major axis. 

   \item Comparison of the {\MgII} kinematics with a disk rotation model for the galaxy shows the {\MgII} spanning a much wider velocity range than the differential rotation of the disk. This implies contribution to the absorption from gas that is not co-rotating with the disk. The ultrastrong {\MgII} absorption is best understood as arising from a combination of circumgalactic structures, some corotating with the disk and the rest at random line-of-sight velocities. The combination of absorption line analysis with IFU data offers a strong means to connect absorbers with the circumgalactic environment of galaxies and infer their origins by comparing them to simulations that model gas around galaxies \citep[e.g.,][]{2021ApJ...923...56D}
   
\end{enumerate}

\begin{acknowledgments}
 SC gratefully acknowledges support from the European Research Council (ERC) under the European Union’s Horizon 2020 Research and Innovation programme grant agreement No 864361
\end{acknowledgments}

\clearpage
\bibliography{ustrong}{}
\bibliographystyle{aasjournal}




\appendix

\renewcommand\thetable{A1}

\begin{minipage}{\textwidth}
    \captionof{table}{Ionization properties of the different components contributing to the absorption with corresponding violin plots in the right-most panel showing posterior distributions for metallicity, hydrogen number density, {\hi} column density, photoionization equilibrium temperature, line of sight thickness, and non-thermal Doppler broadening parameter. The $^{*}$ on top of the violin plot indicates the median value of the distribution. These posterior distributions reflect only the statistical uncertainties. Columns: (1) Velocity of the cloud with respect to the systemic redshift of the galaxy; (2) metallicity of the cloud; (3) total hydrogen number density of the cloud; (4) neutral hydrogen column density of the cloud; (5) temperature of the cloud; (6) inferred line of sight thickness of the cloud; (7) non-thermal Doppler broadening parameter of the cloud (8) thermal Doppler broadening parameter measured for {\hi}; (9) total Doppler broadening parameter measured for {\hi}.}
    \label{tab:cloudproperties}
\end{minipage}

\begin{table*}
        \begin{minipage}[b]{0.4\textwidth}
            \centering
            \begin{sideways} 
                \includegraphics[width=1.0\textwidth,height=2.3\linewidth]{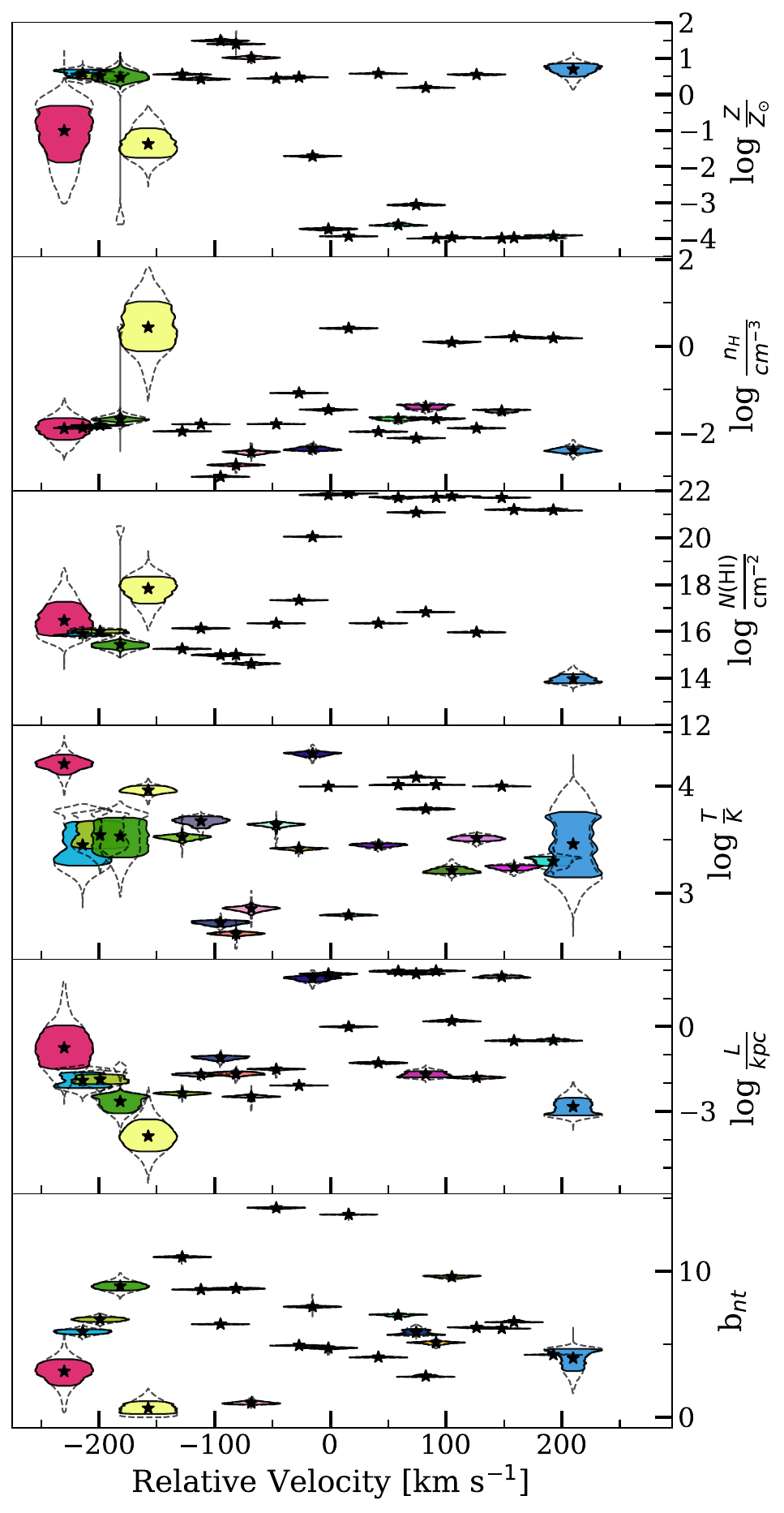}
            \end{sideways}
        \end{minipage}%
        \vfill
        \begin{minipage}[b]{0.65\textwidth}
            \begin{sideways} 
                \begin{tabular}{ccccccccc}
                    \hline\hline
                    (1) & (2) & (3) & (4) & (5) & (6) & (7) & (8) & (9) \\
                    $V$ & \metallicity & \hdenu & \coldenu & \tempu & \thickness & \bturb & \btherm & \bnet \\
                    \kms & & & & & kpc & \kms & \kms & \kms \\
                    \hline
                    \input{tables/cloudproperties.txt}
                \end{tabular}
            \end{sideways}
        \end{minipage}
\end{table*}

\newpage
\renewcommand\thetable{A2}
\input{tables/absorptionproperties.tex}

\pagebreak
\begin{table*}
    \centering
    \renewcommand\thetable{A3}
    \caption{Line measurements of all detected ions-rest frame equivalent width, column density of the velocity spread, and Doppler parameter using the AOD method.}
    \label{tab:A2}
    \begin{tabular}{cccc}
    \toprule
    \multicolumn{4}{c}{AOD measurements}\\
    \hline
    \multicolumn{1}{l}{Line} & \multicolumn{1}{c}{$W_r~(m\angstrom$)} &  \multicolumn{1}{c}{$\log [N_a/{\cmsq}]$} & \multicolumn{1}{c}{v (\kms)} \\ 
    \midrule

 $\MgII~2796$ & $3185.8 \pm 32.2$ & $\gtrsim14.46$ & $[-260,235]$  \\
 $\MgII~2803$ & $2801.7 \pm 28.5$ & $\gtrsim14.63$ & $[-260,235]$  \\
 $\FeII~2344$ & $1301.1\pm17.4$ & $\gtrsim14.78$ & $[-260,235]$ \\
$\MgI~2852$ &  $419.6\pm8.2$ & $12.65\pm0.01$ & $[-260,235]$\\
$\CaII~3934$ & $223.1\pm9.9$ & $12.48\pm0.02$ & $[-170,210]$ \\
$\TiII~3384$ & $14.2\pm2.6$ & $11.60\pm0.07$ & $[-50,5]$\\

    \bottomrule
    \end{tabular}
\end{table*}

\end{document}

%% file: tables/cloudproperties.txt
\colorbox{comp1}{$-230.01_{-0.07}^{+0.15}$} & $-1.01_{-0.99}^{+0.75}$ & $-1.90_{-0.28}^{+0.25}$ & $16.46_{-0.69}^{+0.86}$ & $4.20_{-0.08}^{+0.07}$ & $-0.75_{-0.76}^{+0.79}$ & $3.1_{-0.9}^{+0.8}$ & $16.3_{-1.4}^{+1.3}$ & $16.6_{-1.3}^{+1.2}$\\[2pt]

\colorbox{comp2}{$-214.09_{-0.07}^{+0.10}$} & $0.59_{-0.17}^{+0.10}$ & $-1.88_{-0.02}^{+0.02}$ & $15.90_{-0.11}^{+0.17}$ & $3.45_{-0.20}^{+0.27}$ & $-1.91_{-0.29}^{+0.35}$ & $5.9_{-0.2}^{+0.2}$ & $6.8_{-1.4}^{+2.5}$ & $9.0_{-1.0}^{+1.9}$\\[2pt]

\colorbox{comp3}{$-199.00_{-0.12}^{+0.13}$}& $0.51_{-0.09}^{+0.09}$ & $-1.82_{-0.01}^{+0.01}$ & $16.00_{-0.10}^{+0.10}$ & $3.54_{-0.14}^{+0.16}$ & $-1.87_{-0.19}^{+0.21}$ & $6.7_{-0.2}^{+0.2}$ & $7.6_{-1.1}^{+1.5}$ & $10.1_{-0.8}^{+1.1}$\\[2pt]

\colorbox{comp4}{$-181.61_{-0.15}^{+0.07}$} &  $0.48_{-0.15}^{+0.14}$ & $-1.68_{-0.04}^{+0.07}$ & $15.44_{-0.14}^{+0.16}$ & $3.53_{-0.21}^{+0.19}$ & $-2.65_{-0.36}^{+0.28}$ & $9.0_{-0.3}^{+0.3}$ & $7.5_{-1.6}^{+1.9}$ & $11.7_{-0.9}^{+1.2}$\\[2pt]

\colorbox{comp5}{$-157.45_{-0.10}^{+0.09}$} & $-1.37_{-0.39}^{+0.48}$ & $0.43_{-0.56}^{+0.62}$ & $17.83_{-0.68}^{+0.53}$ & $3.96_{-0.03}^{+0.04}$ & $-3.88_{-0.52}^{+0.58}$ & $0.6_{-0.4}^{+0.5}$ & $12.3_{-0.5}^{+0.5}$ & $12.3_{-0.5}^{+0.5}$\\[2pt]

\colorbox{comp6}{$-128.07_{-0.00}^{+0.00}$} & $0.56_{-0.01}^{+0.01}$ & $-1.96_{-0.01}^{+0.01}$ & $15.26_{-0.05}^{+0.05}$ & $3.52_{-0.02}^{+0.02}$ & $-2.36_{-0.07}^{+0.06}$ & $11.0_{-0.0}^{+0.0}$ & $7.4_{-0.2}^{+0.2}$ & $13.3_{-0.1}^{+0.1}$\\[2pt]

\colorbox{comp7}{$-111.87_{-0.00}^{+0.00}$} & $0.43_{-0.02}^{+0.03}$ & $-1.80_{-0.00}^{+0.00}$ & $16.13_{-0.06}^{+0.05}$ & $3.67_{-0.05}^{+0.03}$ & $-1.70_{-0.07}^{+0.07}$ & $8.8_{-0.0}^{+0.0}$ & $8.8_{-0.5}^{+0.4}$ & $12.4_{-0.3}^{+0.3}$\\[2pt]

\colorbox{comp8}{\textcolor{white}{$-94.97_{-0.00}^{+0.00}$}} & $1.50_{-0.02}^{+0.03}$ & $-3.01_{-0.01}^{+0.02}$ & $14.99_{-0.06}^{+0.06}$ & $2.72_{-0.02}^{+0.02}$ & $-1.10_{-0.10}^{+0.07}$ & $6.4_{-0.0}^{+0.0}$ & $2.9_{-0.1}^{+0.1}$ & $7.0_{-0.0}^{+0.0}$\\[2pt]

\colorbox{comp9}{$-81.59_{-0.02}^{+0.01}$} & $1.40_{-0.01}^{+0.01}$ & $-2.74_{-0.02}^{+0.02}$ & $15.01_{-0.05}^{+0.05}$ & $2.62_{-0.02}^{+0.01}$ & $-1.68_{-0.08}^{+0.08}$ & $8.8_{-0.0}^{+0.0}$ & $2.6_{-0.1}^{+0.0}$ & $9.2_{-0.1}^{+0.1}$\\[2pt]

\colorbox{comp10}{$-68.39_{-0.02}^{+0.02}$} & $1.02_{-0.03}^{+0.03}$ & $-2.44_{-0.04}^{+0.03}$ & $14.62_{-0.06}^{+0.06}$ & $2.85_{-0.03}^{+0.02}$ & $-2.48_{-0.06}^{+0.06}$ & $1.0_{-0.1}^{+0.1}$ & $3.4_{-0.1}^{+0.1}$ & $3.6_{-0.1}^{+0.1}$\\[2pt]

\colorbox{comp11}{$-46.83_{-0.01}^{+0.01}$} & $0.45_{-0.01}^{+0.02}$ & $-1.80_{-0.00}^{+0.00}$ & $16.35_{-0.05}^{+0.05}$ & $3.64_{-0.03}^{+0.01}$ & $-1.51_{-0.06}^{+0.05}$ & $14.4_{-0.0}^{+0.0}$ & $8.4_{-0.3}^{+0.1}$ & $16.7_{-0.1}^{+0.1}$\\[2pt]

\colorbox{comp13}{$-27.16_{-0.04}^{+0.04}$} & $0.48_{-0.01}^{+0.01}$ & $-1.08_{-0.01}^{+0.01}$ & $17.33_{-0.05}^{+0.05}$ & $3.41_{-0.01}^{+0.01}$ & $-2.09_{-0.05}^{+0.05}$ & $4.9_{-0.0}^{+0.0}$ & $6.5_{-0.1}^{+0.1}$ & $8.2_{-0.1}^{+0.0}$\\[2pt]

\colorbox{comp14}{\textcolor{white}{$-15.46_{-0.03}^{+0.07}$}} & $-1.71_{-0.02}^{+0.02}$ & $-2.38_{-0.03}^{+0.06}$ & $20.04_{-0.05}^{+0.05}$ & $4.30_{-0.03}^{+0.02}$ & $1.71_{-0.13}^{+0.09}$ & $7.6_{-0.0}^{+0.1}$ & $18.1_{-0.5}^{+0.4}$ & $19.7_{-0.5}^{+0.4}$\\[2pt]

\colorbox{comp15}{\textcolor{white}{$-1.77_{-0.01}^{+0.01}$}} & $-3.73_{-0.03}^{+0.02}$ & $-1.47_{-0.02}^{+0.01}$ & $21.84_{-0.05}^{+0.06}$ & $3.99_{-0.00}^{+0.00}$ & $1.86_{-0.06}^{+0.06}$ & $4.7_{-0.1}^{+0.0}$ & $12.7_{-0.0}^{+0.0}$ & $13.6_{-0.0}^{+0.0}$\\[2pt]

\colorbox{comp16}{$17.78_{-0.06}^{+0.08}$} & $-3.50_{-0.20}^{+0.14}$ & $0.40_{-0.02}^{+0.02}$ & $21.36_{-0.15}^{+0.20}$ & $2.92_{-0.06}^{+0.04}$ & $-0.53_{-0.15}^{+0.21}$ & $10.9_{-0.1}^{+0.1}$ & $3.7_{-0.2}^{+0.2}$ & $11.5_{-0.2}^{+0.1}$\\[2pt]

\colorbox{comp17}{\textcolor{white}{$41.42_{-0.01}^{+0.01}$}} & $0.58_{-0.01}^{+0.01}$ & $-1.97_{-0.01}^{+0.01}$ & $16.35_{-0.05}^{+0.05}$ & $3.45_{-0.02}^{+0.02}$ & $-1.29_{-0.06}^{+0.06}$ & $4.1_{-0.0}^{+0.0}$ & $6.8_{-0.2}^{+0.1}$ & $7.9_{-0.1}^{+0.1}$\\[2pt]

\colorbox{comp18}{$58.58_{-0.10}^{+0.06}$} & $-3.62_{-0.02}^{+0.05}$ & $-1.68_{-0.05}^{+0.04}$ & $21.72_{-0.07}^{+0.06}$ & $4.01_{-0.00}^{+0.00}$ & $1.96_{-0.06}^{+0.06}$ & $7.0_{-0.1}^{+0.1}$ & $13.0_{-0.0}^{+0.0}$ & $14.7_{-0.0}^{+0.0}$\\[2pt]

\colorbox{comp19}{\textcolor{white}{$74.09_{-0.12}^{+0.08}$}} & $-3.06_{-0.02}^{+0.03}$ & $-2.12_{-0.02}^{+0.01}$ & $21.08_{-0.06}^{+0.05}$ & $4.08_{-0.00}^{+0.01}$ & $1.87_{-0.06}^{+0.06}$ & $5.8_{-0.2}^{+0.3}$ & $14.1_{-0.0}^{+0.1}$ & $15.2_{-0.1}^{+0.2}$\\[2pt]

\colorbox{comp20}{$82.34_{-0.03}^{+0.02}$} & $0.19_{-0.01}^{+0.02}$ & $-1.40_{-0.07}^{+0.05}$ & $16.83_{-0.05}^{+0.05}$ & $3.78_{-0.01}^{+0.01}$ & $-1.69_{-0.12}^{+0.14}$ & $2.8_{-0.1}^{+0.1}$ & $10.0_{-0.1}^{+0.1}$ & $10.4_{-0.1}^{+0.1}$\\[2pt]

\colorbox{comp21}{$91.24_{-0.06}^{+0.14}$} & $-3.99_{-0.00}^{+0.01}$ & $-1.68_{-0.02}^{+0.02}$ & $21.73_{-0.05}^{+0.05}$ & $4.01_{-0.00}^{+0.00}$ & $1.97_{-0.05}^{+0.05}$ & $5.1_{-0.1}^{+0.1}$ & $13.0_{-0.0}^{+0.0}$ & $13.9_{-0.0}^{+0.0}$\\[2pt]

\colorbox{comp22}{$104.90_{-0.07}^{+0.11}$} & $-3.96_{-0.02}^{+0.01}$ & $0.09_{-0.03}^{+0.02}$ & $21.77_{-0.05}^{+0.05}$ & $3.21_{-0.03}^{+0.04}$ & $0.20_{-0.06}^{+0.06}$ & $9.6_{-0.1}^{+0.1}$ & $5.2_{-0.2}^{+0.2}$ & $10.9_{-0.1}^{+0.1}$\\[2pt]

\colorbox{comp23}{$126.21_{-0.01}^{+0.02}$} & $0.55_{-0.02}^{+0.02}$ & $-1.89_{-0.01}^{+0.01}$ & $15.97_{-0.05}^{+0.05}$ & $3.51_{-0.03}^{+0.03}$ & $-1.80_{-0.06}^{+0.06}$ & $6.2_{-0.0}^{+0.0}$ & $7.3_{-0.2}^{+0.2}$ & $9.5_{-0.2}^{+0.2}$\\[2pt]

\colorbox{comp24}{$147.98_{-0.03}^{+0.05}$} & $-3.99_{-0.01}^{+0.02}$ & $-1.50_{-0.04}^{+0.03}$ & $21.71_{-0.05}^{+0.05}$ & $4.00_{-0.00}^{+0.00}$ & $1.77_{-0.07}^{+0.07}$ & $6.1_{-0.0}^{+0.0}$ & $12.8_{-0.0}^{+0.0}$ & $14.2_{-0.0}^{+0.0}$\\[2pt]

\colorbox{comp25}{$158.60_{-0.00}^{+0.00}$} & $-3.97_{-0.02}^{+0.02}$ & $0.21_{-0.01}^{+0.02}$ & $21.20_{-0.05}^{+0.05}$ & $3.24_{-0.03}^{+0.02}$ & $-0.50_{-0.05}^{+0.05}$ & $6.6_{-0.0}^{+0.1}$ & $5.3_{-0.2}^{+0.1}$ & $8.4_{-0.1}^{+0.1}$\\[2pt]

\colorbox{comp26}{$192.58_{-0.01}^{+0.01}$} & $-3.93_{-0.05}^{+0.03}$ & $0.19_{-0.01}^{+0.02}$ & $21.19_{-0.06}^{+0.07}$ & $3.30_{-0.05}^{+0.04}$ & $-0.48_{-0.06}^{+0.07}$ & $4.3_{-0.0}^{+0.0}$ & $5.7_{-0.3}^{+0.3}$ & $7.1_{-0.2}^{+0.2}$\\[2pt]

\colorbox{comp27}{$209.68_{-0.22}^{+0.15}$} & $0.70_{-0.22}^{+0.16}$ & $-2.40_{-0.07}^{+0.09}$ & $13.98_{-0.19}^{+0.23}$ & $3.46_{-0.34}^{+0.35}$ & $-2.83_{-0.33}^{+0.36}$ & $4.1_{-1.0}^{+0.6}$ & $6.9_{-2.2}^{+3.4}$ & $8.0_{-1.5}^{+2.8}$\\
\hline

%% file: tables/absorptionproperties.tex
\small
\addtolength{\tabcolsep}{-1.5pt}
\startlongtable
\begin{deluxetable*}{ccccc}
\tablecaption{Absorption properties of the components estimated by CLOUDY ionization modelling\label{tab:abs-properties}}
\tablehead{
\colhead{Ion} & \colhead{$V$} & \colhead{$b$} & \colhead{$\log N$} & \colhead{$\log \Sigma\,N$}\\
\colhead{} & \colhead{(\kms)} & \colhead{(\kms)} & \colhead{}& \colhead{}
}
\startdata
\hline
\midrule
\input{tables/ColumnDensities.tex}\\
\enddata
 \tablecomments{Absorption properties of the different components contributing to the absorption. Columns: (1) Ion name; (2) Velocity of the cloud with respect to the systemic redshift of the galaxy; (3) total Doppler broadening parameter estimated for the ion; (4) column density of the ion intercepted through the cloud. The quoted 1$\sigma$ uncertainties include both statistical and systematic uncertainties added in quadrature. The statistical uncertainties arise from the noise in the observed data. The systematic uncertainties arise from \textsc{cloudy} modeling which contributes a uniform uncertainty value of 0.05 dex in the predicted column densities. When the absorption is a non-detection at the 3$\sigma$ threshold, we adopt the 2$\sigma$ column density upper limit derived from the equivalent width uncertainty assuming the line lies on the linear curve of growth; (5) total column density of the ion.}
\end{deluxetable*}

%% file: tables/ColumnDensities.tex
CaII & $-230.01_{-0.07}^{+0.15}$ & ${\cdots}$ & $<10.99$ & $13.24_{-0.02}^{+0.02}$\\[2pt]
& $-214.09_{-0.07}^{+0.10}$ & $5.98_{-0.15}^{+0.15}$ & $11.73_{-0.05}^{+0.05}$ & \\[2pt]
& $-199.00_{-0.12}^{+0.13}$ & $6.82_{-0.16}^{+0.16}$ & $11.76_{-0.05}^{+0.05}$ & \\[2pt]
& $-181.61_{-0.15}^{+0.07}$ & ${\cdots}$ & $<10.99$ & \\[2pt]
& $-157.45_{-0.10}^{+0.09}$ & ${\cdots}$ & $<10.99$ & \\[2pt]
& $-128.07_{-0.00}^{+0.00}$ & ${\cdots}$ & $<10.99$ & \\[2pt]
& $-111.87_{-0.00}^{+0.00}$ & $8.87_{-0.03}^{+0.03}$ & $11.81_{-0.05}^{+0.05}$ & \\[2pt]
& $-94.97_{-0.00}^{+0.00}$ & $6.40_{-0.01}^{+0.03}$ & $11.51_{-0.05}^{+0.05}$ & \\[2pt]
& $-81.59_{-0.02}^{+0.01}$ & $8.83_{-0.05}^{+0.05}$ & $11.58_{-0.05}^{+0.05}$ & \\[2pt]
& $-68.39_{-0.02}^{+0.02}$ & ${\cdots}$ & $<10.99$ & \\[2pt]
& $-46.83_{-0.01}^{+0.01}$ & $14.42_{-0.05}^{+0.04}$ & $12.04_{-0.05}^{+0.05}$ & \\[2pt]
& $-27.16_{-0.04}^{+0.04}$ & $5.03_{-0.02}^{+0.03}$ & $12.87_{-0.05}^{+0.05}$ & \\[2pt]
& $-15.46_{-0.03}^{+0.07}$ & $8.12_{-0.06}^{+0.07}$ & $11.53_{-0.05}^{+0.05}$ & \\[2pt]
& $-1.77_{-0.01}^{+0.01}$ & $5.16_{-0.06}^{+0.04}$ & $11.23_{-0.05}^{+0.05}$ & \\[2pt]
& $17.78_{-0.06}^{+0.08}$ & $10.92_{-0.11}^{+0.07}$ & $12.02_{-0.05}^{+0.05}$ & \\[2pt]
& $41.42_{-0.01}^{+0.01}$ & $4.26_{-0.03}^{+0.04}$ & $12.17_{-0.05}^{+0.05}$ & \\[2pt]
& $58.58_{-0.10}^{+0.06}$ & ${\cdots}$ & $<10.99$ & \\[2pt]
& $74.09_{-0.12}^{+0.08}$ & ${\cdots}$ & $<10.99$ & \\[2pt]
& $82.34_{-0.03}^{+0.02}$ & $3.24_{-0.05}^{+0.07}$ & $12.28_{-0.05}^{+0.06}$ & \\[2pt]
& $91.24_{-0.06}^{+0.14}$ & ${\cdots}$ & $<10.99$ & \\[2pt]
& $104.90_{-0.07}^{+0.11}$ & $9.67_{-0.11}^{+0.07}$ & $11.85_{-0.05}^{+0.05}$ & \\[2pt]
& $126.21_{-0.01}^{+0.02}$ & $6.27_{-0.05}^{+0.03}$ & $11.75_{-0.05}^{+0.05}$ & \\[2pt]
& $147.98_{-0.03}^{+0.05}$ & ${\cdots}$ & $<10.99$ & \\[2pt]
& $158.60_{-0.00}^{+0.00}$ & $6.61_{-0.04}^{+0.07}$ & $11.32_{-0.05}^{+0.05}$ & \\[2pt]
& $192.58_{-0.01}^{+0.01}$ & $4.39_{-0.02}^{+0.02}$ & $11.33_{-0.05}^{+0.05}$ & \\[2pt]
& $209.68_{-0.22}^{+0.15}$ & ${\cdots}$ & $<10.99$ & \\[2pt]
\hline
FeII & $-230.01_{-0.07}^{+0.15}$ & ${\cdots}$ & $<11.59$ & $14.72_{-0.02}^{+0.02}$\\[2pt]
& $-214.09_{-0.07}^{+0.10}$ & $5.95_{-0.16}^{+0.15}$ & $12.68_{-0.05}^{+0.05}$ & \\[2pt]
& $-199.00_{-0.12}^{+0.13}$ & $6.79_{-0.16}^{+0.16}$ & $12.81_{-0.05}^{+0.05}$ & \\[2pt]
& $-181.61_{-0.15}^{+0.07}$ & $9.05_{-0.30}^{+0.30}$ & $12.35_{-0.06}^{+0.06}$ & \\[2pt]
& $-157.45_{-0.10}^{+0.09}$ & $1.78_{-0.12}^{+0.22}$ & $12.14_{-0.05}^{+0.05}$ & \\[2pt]
& $-128.07_{-0.00}^{+0.00}$ & $11.04_{-0.04}^{+0.04}$ & $11.90_{-0.05}^{+0.05}$ & \\[2pt]
& $-111.87_{-0.00}^{+0.00}$ & $8.84_{-0.03}^{+0.03}$ & $12.90_{-0.05}^{+0.05}$ & \\[2pt]
& $-94.97_{-0.00}^{+0.00}$ & ${\cdots}$ & $<11.59$ & \\[2pt]
& $-81.59_{-0.02}^{+0.01}$ & ${\cdots}$ & $<11.59$ & \\[2pt]
& $-68.39_{-0.02}^{+0.02}$ & ${\cdots}$ & $<11.59$ & \\[2pt]
& $-46.83_{-0.01}^{+0.01}$ & $14.41_{-0.05}^{+0.04}$ & $13.13_{-0.05}^{+0.05}$ & \\[2pt]
& $-27.16_{-0.04}^{+0.04}$ & $5.00_{-0.02}^{+0.03}$ & $14.20_{-0.05}^{+0.05}$ & \\[2pt]
& $-15.46_{-0.03}^{+0.07}$ & $7.97_{-0.05}^{+0.06}$ & $13.91_{-0.05}^{+0.05}$ & \\[2pt]
& $-1.77_{-0.01}^{+0.01}$ & $5.05_{-0.07}^{+0.04}$ & $13.59_{-0.05}^{+0.05}$ & \\[2pt]
& $17.78_{-0.06}^{+0.08}$ & $10.91_{-0.11}^{+0.07}$ & $13.31_{-0.05}^{+0.05}$ & \\[2pt]
& $41.42_{-0.01}^{+0.01}$ & $4.22_{-0.03}^{+0.04}$ & $12.99_{-0.05}^{+0.05}$ & \\[2pt]
& $58.58_{-0.10}^{+0.06}$ & $7.22_{-0.11}^{+0.06}$ & $13.60_{-0.05}^{+0.05}$ & \\[2pt]
& $74.09_{-0.12}^{+0.08}$ & $6.13_{-0.18}^{+0.27}$ & $13.55_{-0.05}^{+0.05}$ & \\[2pt]
& $82.34_{-0.03}^{+0.02}$ & $3.13_{-0.05}^{+0.08}$ & $13.62_{-0.05}^{+0.05}$ & \\[2pt]
& $91.24_{-0.06}^{+0.14}$ & $5.40_{-0.13}^{+0.08}$ & $13.23_{-0.05}^{+0.05}$ & \\[2pt]
& $104.90_{-0.07}^{+0.11}$ & $9.66_{-0.11}^{+0.07}$ & $13.27_{-0.05}^{+0.05}$ & \\[2pt]
& $126.21_{-0.01}^{+0.02}$ & $6.24_{-0.05}^{+0.03}$ & $12.71_{-0.05}^{+0.05}$ & \\[2pt]
& $147.98_{-0.03}^{+0.05}$ & $6.32_{-0.01}^{+0.02}$ & $13.22_{-0.05}^{+0.05}$ & \\[2pt]
& $158.60_{-0.00}^{+0.00}$ & $6.59_{-0.04}^{+0.07}$ & $12.69_{-0.05}^{+0.05}$ & \\[2pt]
& $192.58_{-0.01}^{+0.01}$ & $4.36_{-0.02}^{+0.02}$ & $12.71_{-0.05}^{+0.05}$ & \\[2pt]
& $209.68_{-0.22}^{+0.15}$ & ${\cdots}$ & $<11.59$ & \\[2pt]
\hline
HI & $-230.01_{-0.07}^{+0.15}$ & $16.6_{-1.3}^{+1.2}$ & $16.46_{-0.69}^{+0.86}$ & $22.55_{-0.02}^{+0.02}$\\[2pt]
& $-214.09_{-0.07}^{+0.10}$ & $9.00_{-0.98}^{+1.91}$ & $15.90_{-0.11}^{+0.17}$ & \\[2pt]
& $-199.00_{-0.12}^{+0.13}$ & $10.12_{-0.78}^{+1.13}$ & $16.00_{-0.10}^{+0.10}$ & \\[2pt]
& $-181.61_{-0.15}^{+0.07}$ & $11.73_{-0.95}^{+1.18}$ & $15.44_{-0.14}^{+0.16}$ & \\[2pt]
& $-157.45_{-0.10}^{+0.09}$ & $12.29_{-0.46}^{+0.50}$ & $17.83_{-0.68}^{+0.53}$ & \\[2pt]
& $-128.07_{-0.00}^{+0.00}$ & $13.25_{-0.14}^{+0.13}$ & $15.26_{-0.05}^{+0.05}$ & \\[2pt]
& $-111.87_{-0.00}^{+0.00}$ & $12.41_{-0.33}^{+0.27}$ & $16.13_{-0.06}^{+0.05}$ & \\[2pt]
& $-94.97_{-0.00}^{+0.00}$ & $7.04_{-0.02}^{+0.02}$ & $14.99_{-0.06}^{+0.06}$ & \\[2pt]
& $-81.59_{-0.02}^{+0.01}$ & $9.20_{-0.06}^{+0.06}$ & $15.01_{-0.05}^{+0.05}$ & \\[2pt]
& $-68.39_{-0.02}^{+0.02}$ & $3.58_{-0.09}^{+0.07}$ & $14.62_{-0.06}^{+0.06}$ & \\[2pt]
& $-46.83_{-0.01}^{+0.01}$ & $16.66_{-0.13}^{+0.06}$ & $16.35_{-0.05}^{+0.05}$ & \\[2pt]
& $-27.16_{-0.04}^{+0.04}$ & $8.18_{-0.06}^{+0.05}$ & $17.33_{-0.05}^{+0.05}$ & \\[2pt]
& $-15.46_{-0.03}^{+0.07}$ & $19.68_{-0.50}^{+0.36}$ & $20.04_{-0.05}^{+0.05}$ & \\[2pt]
& $-1.77_{-0.01}^{+0.01}$ & $13.60_{-0.01}^{+0.01}$ & $21.84_{-0.05}^{+0.06}$ & \\[2pt]
& $17.78_{-0.06}^{+0.08}$ & $11.51_{-0.15}^{+0.10}$ & $21.36_{-0.15}^{+0.20}$ & \\[2pt]
& $41.42_{-0.01}^{+0.01}$ & $7.93_{-0.14}^{+0.12}$ & $16.35_{-0.05}^{+0.05}$ & \\[2pt]
& $58.58_{-0.10}^{+0.06}$ & $14.74_{-0.03}^{+0.03}$ & $21.72_{-0.07}^{+0.06}$ & \\[2pt]
& $74.09_{-0.12}^{+0.08}$ & $15.24_{-0.10}^{+0.17}$ & $21.08_{-0.06}^{+0.05}$ & \\[2pt]
& $82.34_{-0.03}^{+0.02}$ & $10.41_{-0.09}^{+0.09}$ & $16.83_{-0.05}^{+0.05}$ & \\[2pt]
& $91.24_{-0.06}^{+0.14}$ & $13.94_{-0.04}^{+0.03}$ & $21.73_{-0.05}^{+0.05}$ & \\[2pt]
& $104.90_{-0.07}^{+0.11}$ & $10.94_{-0.09}^{+0.10}$ & $21.77_{-0.05}^{+0.05}$ & \\[2pt]
& $126.21_{-0.01}^{+0.02}$ & $9.54_{-0.17}^{+0.18}$ & $15.97_{-0.05}^{+0.05}$ & \\[2pt]
& $147.98_{-0.03}^{+0.05}$ & $14.16_{-0.03}^{+0.05}$ & $21.71_{-0.05}^{+0.05}$ & \\[2pt]
& $158.60_{-0.00}^{+0.00}$ & $8.45_{-0.09}^{+0.07}$ & $21.20_{-0.05}^{+0.05}$ & \\[2pt]
& $192.58_{-0.01}^{+0.01}$ & $7.14_{-0.25}^{+0.22}$ & $21.19_{-0.06}^{+0.07}$ & \\[2pt]
& $209.68_{-0.22}^{+0.15}$ & $7.99_{-1.46}^{+2.76}$ & $13.98_{-0.19}^{+0.23}$ & \\[2pt]
\hline
MgI & $-230.01_{-0.07}^{+0.15}$ & ${\cdots}$ & $<10.55$ & $12.69_{-0.02}^{+0.02}$\\[2pt]
& $-214.09_{-0.07}^{+0.10}$ & ${\cdots}$ & $<10.55$ & \\[2pt]
& $-199.00_{-0.12}^{+0.13}$ & ${\cdots}$ & $<10.55$ & \\[2pt]
& $-181.61_{-0.15}^{+0.07}$ & ${\cdots}$ & $<10.55$ & \\[2pt]
& $-157.45_{-0.10}^{+0.09}$ & $2.60_{-0.12}^{+0.15}$ & $10.93_{-0.06}^{+0.06}$ & \\[2pt]
& $-128.07_{-0.00}^{+0.00}$ & ${\cdots}$ & $<10.55$ & \\[2pt]
& $-111.87_{-0.00}^{+0.00}$ & ${\cdots}$ & $<10.55$ & \\[2pt]
& $-94.97_{-0.00}^{+0.00}$ & ${\cdots}$ & $<10.55$ & \\[2pt]
& $-81.59_{-0.02}^{+0.01}$ & ${\cdots}$ & $<10.55$ & \\[2pt]
& $-68.39_{-0.02}^{+0.02}$ & ${\cdots}$ & $<10.55$ & \\[2pt]
& $-46.83_{-0.01}^{+0.01}$ & $14.46_{-0.05}^{+0.04}$ & $10.97_{-0.05}^{+0.05}$ & \\[2pt]
& $-27.16_{-0.04}^{+0.04}$ & $5.10_{-0.02}^{+0.03}$ & $12.35_{-0.05}^{+0.05}$ & \\[2pt]
& $-15.46_{-0.03}^{+0.07}$ & $8.44_{-0.07}^{+0.08}$ & $11.71_{-0.05}^{+0.05}$ & \\[2pt]
& $-1.77_{-0.01}^{+0.01}$ & $5.41_{-0.06}^{+0.04}$ & $11.52_{-0.05}^{+0.05}$ & \\[2pt]
& $17.78_{-0.06}^{+0.08}$ & $10.93_{-0.11}^{+0.07}$ & $11.40_{-0.05}^{+0.05}$ & \\[2pt]
& $41.42_{-0.01}^{+0.01}$ & $4.35_{-0.03}^{+0.04}$ & $11.00_{-0.05}^{+0.05}$ & \\[2pt]
& $58.58_{-0.10}^{+0.06}$ & $7.49_{-0.10}^{+0.06}$ & $11.45_{-0.05}^{+0.05}$ & \\[2pt]
& $74.09_{-0.12}^{+0.08}$ & $6.49_{-0.17}^{+0.26}$ & $11.21_{-0.05}^{+0.05}$ & \\[2pt]
& $82.34_{-0.03}^{+0.02}$ & $3.49_{-0.05}^{+0.07}$ & $11.59_{-0.05}^{+0.05}$ & \\[2pt]
& $91.24_{-0.06}^{+0.14}$ & $5.75_{-0.12}^{+0.08}$ & $11.09_{-0.05}^{+0.05}$ & \\[2pt]
& $104.90_{-0.07}^{+0.11}$ & $9.70_{-0.11}^{+0.07}$ & $11.19_{-0.05}^{+0.05}$ & \\[2pt]
& $126.21_{-0.01}^{+0.02}$ & ${\cdots}$ & $<10.55$ & \\[2pt]
& $147.98_{-0.03}^{+0.05}$ & $6.62_{-0.01}^{+0.02}$ & $11.14_{-0.05}^{+0.05}$ & \\[2pt]
& $158.60_{-0.00}^{+0.00}$ & ${\cdots}$ & $<10.55$ & \\[2pt]
& $192.58_{-0.01}^{+0.01}$ & ${\cdots}$ & $<10.55$ & \\[2pt]
& $209.68_{-0.22}^{+0.15}$ & ${\cdots}$ & $<10.55$ & \\[2pt]
\hline
MgII & $-230.01_{-0.07}^{+0.15}$ & $4.6_{-0.5}^{+0.5}$ & $11.97_{-0.06}^{+0.06}$ & $14.76_{-0.02}^{+0.02}$\\[2pt]
& $-214.09_{-0.07}^{+0.10}$ & $6.05_{-0.15}^{+0.14}$ & $13.07_{-0.05}^{+0.05}$ & \\[2pt]
& $-199.00_{-0.12}^{+0.13}$ & $6.89_{-0.15}^{+0.16}$ & $13.09_{-0.05}^{+0.05}$ & \\[2pt]
& $-181.61_{-0.15}^{+0.07}$ & $9.13_{-0.29}^{+0.29}$ & $12.46_{-0.05}^{+0.05}$ & \\[2pt]
& $-157.45_{-0.10}^{+0.09}$ & $2.60_{-0.12}^{+0.15}$ & $12.19_{-0.05}^{+0.05}$ & \\[2pt]
& $-128.07_{-0.00}^{+0.00}$ & $11.09_{-0.05}^{+0.04}$ & $12.39_{-0.05}^{+0.05}$ & \\[2pt]
& $-111.87_{-0.00}^{+0.00}$ & $8.94_{-0.04}^{+0.04}$ & $13.14_{-0.05}^{+0.05}$ & \\[2pt]
& $-94.97_{-0.00}^{+0.00}$ & $6.42_{-0.01}^{+0.03}$ & $12.94_{-0.05}^{+0.05}$ & \\[2pt]
& $-81.59_{-0.02}^{+0.01}$ & $8.84_{-0.05}^{+0.05}$ & $13.00_{-0.05}^{+0.05}$ & \\[2pt]
& $-68.39_{-0.02}^{+0.02}$ & $1.20_{-0.06}^{+0.11}$ & $12.24_{-0.06}^{+0.05}$ & \\[2pt]
& $-46.83_{-0.01}^{+0.01}$ & $14.46_{-0.05}^{+0.04}$ & $13.38_{-0.05}^{+0.05}$ & \\[2pt]
& $-27.16_{-0.04}^{+0.04}$ & $5.10_{-0.02}^{+0.03}$ & $14.12_{-0.05}^{+0.05}$ & \\[2pt]
& $-15.46_{-0.03}^{+0.07}$ & $8.44_{-0.07}^{+0.08}$ & $13.76_{-0.06}^{+0.06}$ & \\[2pt]
& $-1.77_{-0.01}^{+0.01}$ & $5.41_{-0.06}^{+0.04}$ & $13.65_{-0.05}^{+0.05}$ & \\[2pt]
& $17.78_{-0.06}^{+0.08}$ & $10.93_{-0.11}^{+0.07}$ & $13.39_{-0.05}^{+0.05}$ & \\[2pt]
& $41.42_{-0.01}^{+0.01}$ & $4.35_{-0.03}^{+0.04}$ & $13.54_{-0.05}^{+0.05}$ & \\[2pt]
& $58.58_{-0.10}^{+0.06}$ & $7.49_{-0.10}^{+0.06}$ & $13.64_{-0.05}^{+0.05}$ & \\[2pt]
& $74.09_{-0.12}^{+0.08}$ & $6.49_{-0.17}^{+0.26}$ & $13.48_{-0.05}^{+0.05}$ & \\[2pt]
& $82.34_{-0.03}^{+0.02}$ & $3.49_{-0.05}^{+0.07}$ & $13.56_{-0.06}^{+0.06}$ & \\[2pt]
& $91.24_{-0.06}^{+0.14}$ & $5.75_{-0.12}^{+0.08}$ & $13.28_{-0.05}^{+0.05}$ & \\[2pt]
& $104.90_{-0.07}^{+0.11}$ & $9.70_{-0.11}^{+0.07}$ & $13.35_{-0.05}^{+0.05}$ & \\[2pt]
& $126.21_{-0.01}^{+0.02}$ & $6.33_{-0.05}^{+0.03}$ & $13.09_{-0.05}^{+0.05}$ & \\[2pt]
& $147.98_{-0.03}^{+0.05}$ & $6.62_{-0.01}^{+0.02}$ & $13.27_{-0.05}^{+0.05}$ & \\[2pt]
& $158.60_{-0.00}^{+0.00}$ & $6.64_{-0.03}^{+0.06}$ & $12.77_{-0.05}^{+0.05}$ & \\[2pt]
& $192.58_{-0.01}^{+0.01}$ & $4.45_{-0.02}^{+0.02}$ & $12.79_{-0.05}^{+0.05}$ & \\[2pt]
& $209.68_{-0.22}^{+0.15}$ & ${\cdots}$ & $<11.10$ & \\[2pt]
\hline
TiII & $-230.01_{-0.07}^{+0.15}$ & ${\cdots}$ & $<11.24$ & $11.56_{-0.05}^{+0.05}$\\[2pt]
& $-214.09_{-0.07}^{+0.10}$ & ${\cdots}$ & $<11.24$ & \\[2pt]
& $-199.00_{-0.12}^{+0.13}$ & ${\cdots}$ & $<11.24$ & \\[2pt]
& $-181.61_{-0.15}^{+0.07}$ & ${\cdots}$ & $<11.24$ & \\[2pt]
& $-157.45_{-0.10}^{+0.09}$ & ${\cdots}$ & $<11.24$ & \\[2pt]
& $-128.07_{-0.00}^{+0.00}$ & ${\cdots}$ & $<11.24$ & \\[2pt]
& $-111.87_{-0.00}^{+0.00}$ & ${\cdots}$ & $<11.24$ & \\[2pt]
& $-94.97_{-0.00}^{+0.00}$ & ${\cdots}$ & $<11.24$ & \\[2pt]
& $-81.59_{-0.02}^{+0.01}$ & ${\cdots}$ & $<11.24$ & \\[2pt]
& $-68.39_{-0.02}^{+0.02}$ & ${\cdots}$ & $<11.24$ & \\[2pt]
& $-46.83_{-0.01}^{+0.01}$ & ${\cdots}$ & $<11.24$ & \\[2pt]
& $-27.16_{-0.04}^{+0.04}$ & $5.01_{-0.02}^{+0.03}$ & $11.56_{-0.05}^{+0.05}$ & \\[2pt]
& $-15.46_{-0.03}^{+0.07}$ & ${\cdots}$ & $<11.24$ & \\[2pt]
& $-1.77_{-0.01}^{+0.01}$ & ${\cdots}$ & $<11.24$ & \\[2pt]
& $17.78_{-0.06}^{+0.08}$ & ${\cdots}$ & $<11.24$ & \\[2pt]
& $41.42_{-0.01}^{+0.01}$ & ${\cdots}$ & $<11.24$ & \\[2pt]
& $58.58_{-0.10}^{+0.06}$ & ${\cdots}$ & $<11.24$ & \\[2pt]
& $74.09_{-0.12}^{+0.08}$ & ${\cdots}$ & $<11.24$ & \\[2pt]
& $82.34_{-0.03}^{+0.02}$ & ${\cdots}$ & $<11.24$ & \\[2pt]
& $91.24_{-0.06}^{+0.14}$ & ${\cdots}$ & $<11.24$ & \\[2pt]
& $104.90_{-0.07}^{+0.11}$ & ${\cdots}$ & $<11.24$ & \\[2pt]
& $126.21_{-0.01}^{+0.02}$ & ${\cdots}$ & $<11.24$ & \\[2pt]
& $147.98_{-0.03}^{+0.05}$ & ${\cdots}$ & $<11.24$ & \\[2pt]
& $158.60_{-0.00}^{+0.00}$ & ${\cdots}$ & $<11.24$ & \\[2pt]
& $192.58_{-0.01}^{+0.01}$ & ${\cdots}$ & $<11.24$ & \\[2pt]
& $209.68_{-0.22}^{+0.15}$ & ${\cdots}$ & $<11.24$ & \\[2pt]
\hline